\documentclass[journal]{IEEEtran}
\IEEEoverridecommandlockouts
\usepackage{cite}
\usepackage{amsmath,amssymb,amsfonts}
\usepackage{algorithmic}
\usepackage{graphicx}
\usepackage{subfigure}
\usepackage{textcomp}
\usepackage{xcolor}
\usepackage{makecell}
\usepackage{multirow}
\usepackage{xspace}
\usepackage{url}
\usepackage{graphicx}
\usepackage[ruled,linesnumbered]{algorithm2e}
\usepackage{subcaption}

\def\BibTeX{{\rm B\kern-.05em{\sc i\kern-.025em b}\kern-.08em
    T\kern-.1667em\lower.7ex\hbox{E}\kern-.125emX}}

\newcommand{\SystemName}{Trident\xspace}
\newcommand{\naxin}[1]{\textcolor{black}{#1}}

\newcommand{\add}[1]{\textcolor{black}{#1}}
\newcommand{\rev}[1]{\textcolor{black}{#1}}
\newcommand{\addsec}{\color{black}}
\newcommand{\water}{\vspace{1.5mm}}
\newcommand{\narrowfig}{\vspace{0mm}}

\begin{document}
\title{\SystemName: Interference Avoidance in Multi-reader Backscatter Network via Frequency-space Division}

\markboth{IEEE/ACM TRANSACTIONS ON NETWORKING}{}

%

\author{Yang Zou, \IEEEmembership{Student Member, IEEE,}
        Xin Na, \IEEEmembership{Student Member, IEEE,}
        Yimiao Sun, \IEEEmembership{Student Member, IEEE,}
        and Yuan He\textsuperscript{*}, \IEEEmembership{Senior Member, IEEE}

%
%
\thanks{Yang Zou, Xin Na, Yimiao Sun and Yuan He are with Tsinghua University, P.R. China.}
\thanks{Email: zouy23@mails.tsinghua.edu.cn, xn20@mails.tsinghua.edu.cn, sym21@mails.tsinghua.edu.cn and heyuan@tsinghua.edu.cn}
\thanks{Yuan He\textsuperscript{*} is the corresponding author}
}

\maketitle

\begin{abstract}
    Backscatter is a key technology for battery-free sensing in industrial IoT applications. To fully cover numerous tags in the deployment area, one often needs to deploy multiple readers, each of which communicates with tags within its communication range. However, the actual backscattered signals from a tag are likely to reach a reader outside its communication range and cause interference. Conventional TDMA or CSMA based approaches for interference avoidance separate readers' media access in time, leading to limited network throughput. In this paper, we propose \SystemName, a novel backscatter design that enables interference avoidance via frequency-space division. By incorporating a tunable bandpass filter and multiple terminal loads, a \SystemName tag can detect its channel condition and adaptively adjust the frequency and the power of its backscattered signals. We further propose a frequency assignment algorithm for the readers. With these designs, all the readers in the network can operate concurrently without being interfered. We implement \SystemName and evaluate its performance under various settings. The results demonstrate that \SystemName enhances the network throughput by 3.18×, compared to the TDMA-based scheme.
\end{abstract}


\IEEEpeerreviewmaketitle

\section{Introduction}

\begin{figure}[t]
    \centering
    \includegraphics[width=0.7\linewidth]{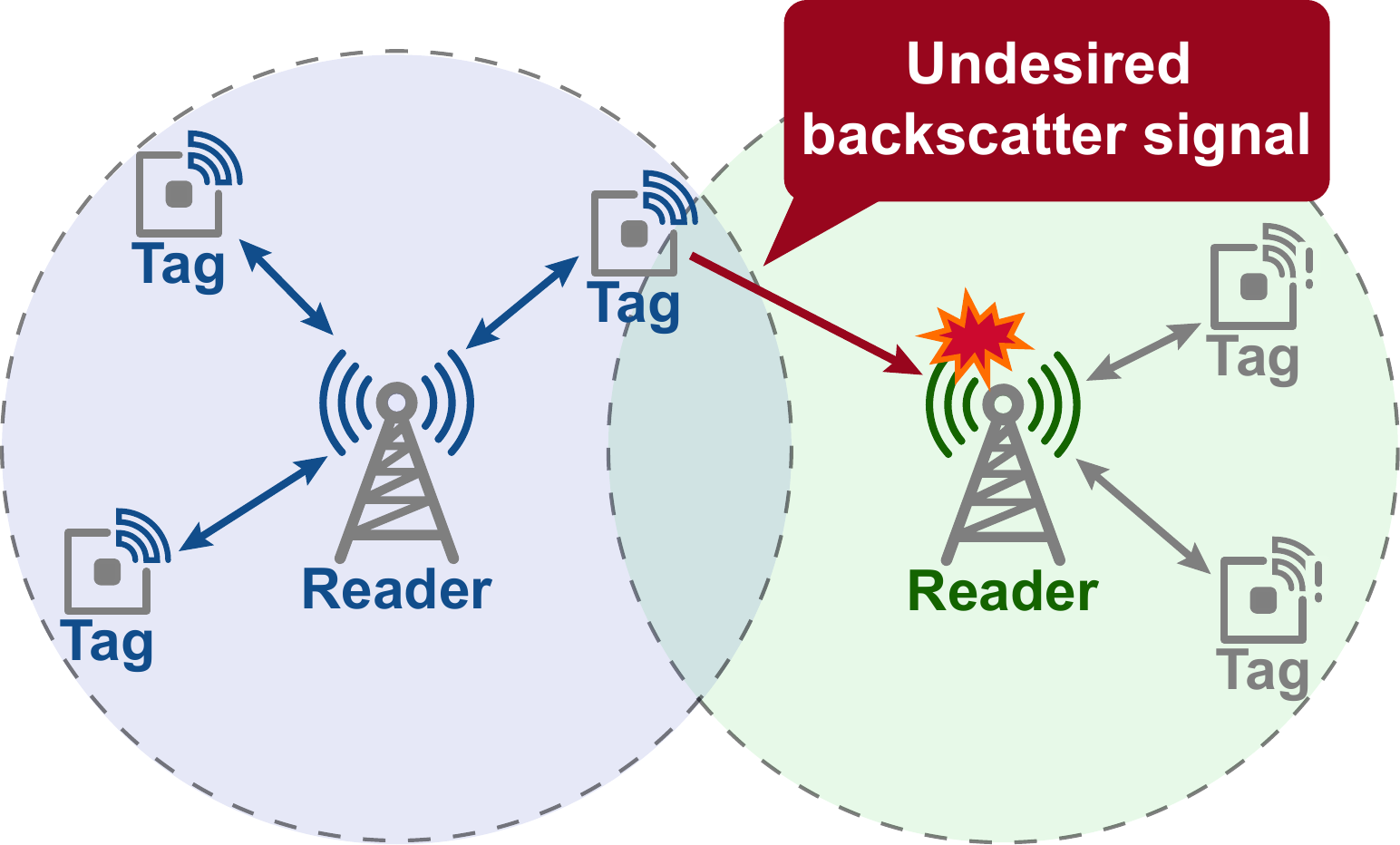}
    \caption{Illustration of interference in the multi-reader backscatter network. }
    \label{fig:interference}
    \narrowfig
    \vspace{-5mm}
\end{figure}

Backscatter is an enabling technology for battery-free sensing applications in the industrial Internet of Things (IoT) \cite{na2023leggiero, talla2017lora, sun2022bifrost, chi2020ltebackscatter, bharadia2015backfi, majid2019multi, kellogg2016passive, abedi2020witag, guo2020aloba, zhang2016enabling, rostami2020redefining}. Deploying a backscatter network in large factories requires dense deployment of readers for full backscatter coverage of the deployment area. In an ideal scenario, each reader covers a small area, enabling excitation and communication with tags in the vicinity. In reality, however, the backscattered signals from a tag are likely to reach a reader which the tag is not intended to communicate with\cite{bletsas2009anti}, inducing undesired interference and potentially degrading the network throughput, as shown in Fig. \ref{fig:interference}.

Conventional approaches to deal with the above interference issue in a backscatter network are either TDMA-based or CSMA-based. In TDMA-based approaches, the readers, need to be carefully coordinated so that adjacent readers do not excite a tag simultaneously \cite{waldrop2003colorwave, gandino2011increasing, zhou2007slotted}. This coordination, whether distributed or centralized, introduces significant complexity and cost in communication. In CSMA-based schemes, a reader needs to take back-offs when the channel is occupied by another reader, and can’t send out the excitation until the channel is clear \cite{birari2005mitigating}. Both TDMA or CSMA based schemes can avoid interference, however, at the cost of network throughput. This is because they sorely focus on the readers and avoid interference only in the time domain.

\rev{In this paper, we consider a novel interference avoidance method, which utilizes other multiplexing spaces to tackle the interference problem and in turn to improve the overall throughput of the backscatter network.}
To achieve this, we propose to empower the backscatter tags with the ability to avoid interference in the frequency and space domains. The backscatter network operates at multiple frequency bands. Adjacent readers are assigned to work at different bands. A tag detects its channel condition and adaptively determines a reader to communicate with. To avoid interference to the readers, the tag should be able to select the frequency band with the strongest excitation signal and only backscatter signals at that band. The power of the backscattered signal should also be finely controlled to avoid interference with another remote reader that works at the same band.

Implementing the above idea is a daunting task, which may meet the following critical challenges:

\begin{enumerate}

    \item Given the highly limited power budget of battery-free tags, typically less than 1mW \cite{dehbashi2021verification}, a significant challenge arises in realizing the aforementioned innovative capabilities involving band selection and frequency-selective reflection.


    \item 
    Selective reflection is a missing piece in the state of the arts. 
    In the existing works, a tag usually backscatters signals across the whole frequency range. How to generate backscattered signals in only one of the multiple bands is a hard problem, especially considering the limited process capacity of a tag.

    \item For the purpose of space division, the power of the backscattered signals should be finely controlled, but how to ensure the efficacy of power control under dynamic channel conditions remains a challenging issue.
    
\end{enumerate}

In this paper, we present \textbf{\SystemName}
, a novel backscatter tag to tackle the aforementioned challenges. A \SystemName tag contains a \textbf{frequency band detector}, which utilizes a frequency-tunable bandpass filter to extract signals from different bands and a low-power two-step comparator for signal strength comparison. We exploit the characteristics of bandpass filters in signal reflection and propose a new scheme for \textbf{frequency-selective reflector}. We further develop a \textbf{reflection power adjuster}, which selects suitable terminal loads for the reflector to control the backscattered signal strength, so that the interference range of the tag doesn't cover the other readers. 
\add{To facilitate the \SystemName's deployment in multi-reader environments, we propose a frequency assignment algorithm, ensuring adjacent readers operate on different bands.}
Including the above designs, \SystemName realizes frequency-space division in a backscatter network and significantly enhances the efficiency of interference avoidance.

Our contributions can be summarized as follows:

\begin{itemize}

    \item \SystemName is the first-of-its-kind work that empowers a backscatter tag with the ability of band detection and frequency-selective reflection in the sub-6G band. 

    \item We propose a reflection power adjustment scheme based on excessive power detection and terminal load selection, which controls the strength of backscattered signals and avoids interfering the same-band readers via space division.

    \item \add{We introduce a frequency assignment algorithm, which assigns frequencies to readers in complex multi-reader deployment, avoiding co-frequency interference among adjacent readers.}
    
    \item We implement \SystemName tag on a printed circuit board (PCB) and deploy a multi-reader backscatter network prototype. The results show that \SystemName enhances the network throughput by 3.18$\times$, compared to the TDMA-based scheme.
    
\end{itemize}

\add{The remainder of this paper is structured as follows. We elaborate on our tag design in Section \ref{title:tag-design} and the frequency assignment algorithm in Section \ref{title:reader-frequency-assignment}. Then we introduce some practical issues in Section \ref{title:practical-issue} and the system implementation in Section \ref{title:implementation}. Section \ref{title:evaluation} presents the evaluation results. Section \ref{title:related-work} reviews the related works. We conclude this paper in Section \ref{title:conclusion}.
}
\section{\SystemName Tag Design}
\label{title:tag-design}
    
\subsection{Overview}

\begin{figure}[t]
    \centering
    \includegraphics[width=0.75\linewidth]{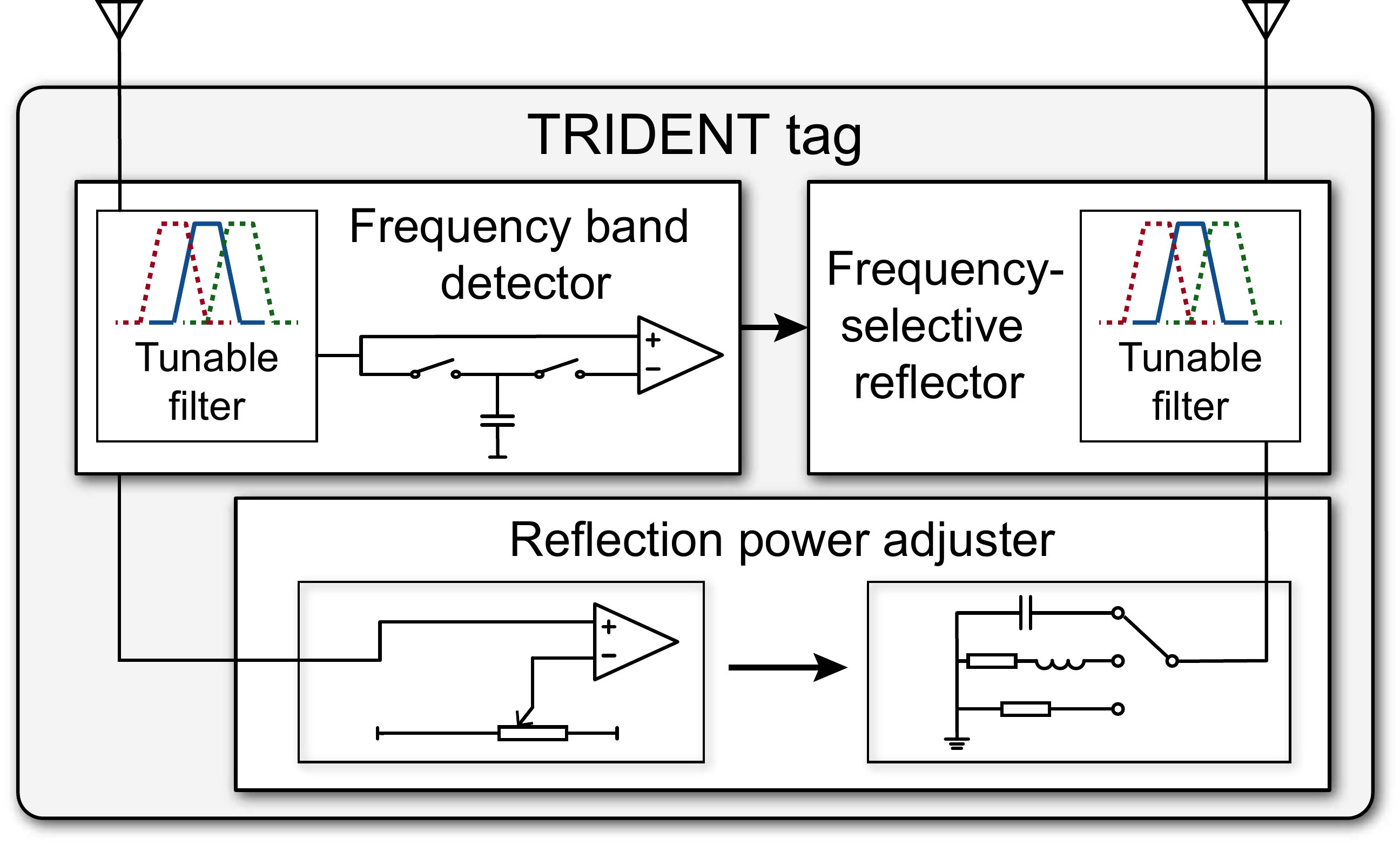}
    \caption{
    The three key components of \SystemName tag.}
    \label{fig:tag-struct-work-flow}
    \narrowfig
    \vspace{-5mm}
\end{figure}

In this section, we propose the design details of \SystemName tag, as shown in Fig. \ref{fig:tag-struct-work-flow}. \SystemName tag mainly consists of three essential components: 

\textit{Frequency band detector:} The frequency band detector accomplishes band detection through two steps: extracting excitation signals at various frequency bands and comparing the strengths pairwise. To extract signals at various bands, we explore a frequency-tunable bandpass filter model that demonstrates the relationship between the center frequency and resonant capacitance, then we implement this filter using varactor diodes. For strength comparison, we notice the signal strength can be converted into voltage by an envelope detector and stored in a capacitor. Thus we utilize the tunable filter to sequentially extract signals from two bands, storing the detection voltage of the previous one and comparing it with the next one. This approach enables low-power strength comparison.

\textit{Frequency-selective reflector:} 
We achieve generating signals only at a specific band by greatly attenuating backscattered signals out of this band. 
For the methodology of generating backscattered signals with different strengths in various bands, we investigate the mathematical model of backscattered signals and identify that the variation of the reflection coefficient determines the backscattered signal strength. Then we find that the bandpass filter's reflection coefficients for signals outside the passband are nearly constant. And based on the filter we develop a frequency-selective reflector.

\textit{Reflection power adjuster:} When the excitation signal is strong, the reflected signal tends to be strong and can cause interference to distant readers operating at the same frequency. To finely control the strength of the backscattered signal, we introduce a detector to detect excessively strong excitation signals based on a threshold voltage comparator. 
Then, building upon the understanding that variations in the reflection coefficient directly impact the strength of backscattered signals, we explore the relationship between the reflection coefficient and the impedance of terminal loads. We design multiple selectable terminal loads that enable adjustment over the backscattered signal strength. Thus we build a reflection power adjuster, enabling adaptively control of the reflection strength. 




Next, we will introduce the frequency band detector and the frequency-selective reflector, followed by the design of the reflection power adjuster.

\subsection{Frequency Band Detector}
\label{title:frequency-band-detector}


\water\subsubsection{Signal Extraction}

The first step in detecting the frequency band with the strongest excitation is to extract excitation signals at each individual band. 
To accomplish this objective while minimizing power consumption, we incorporat a frequency-tunable bandpass filter in the next stage of the antenna\cite{hunter1982tunablefilter}. 

\begin{figure}
    \centering
    \includegraphics[width=0.8\linewidth]{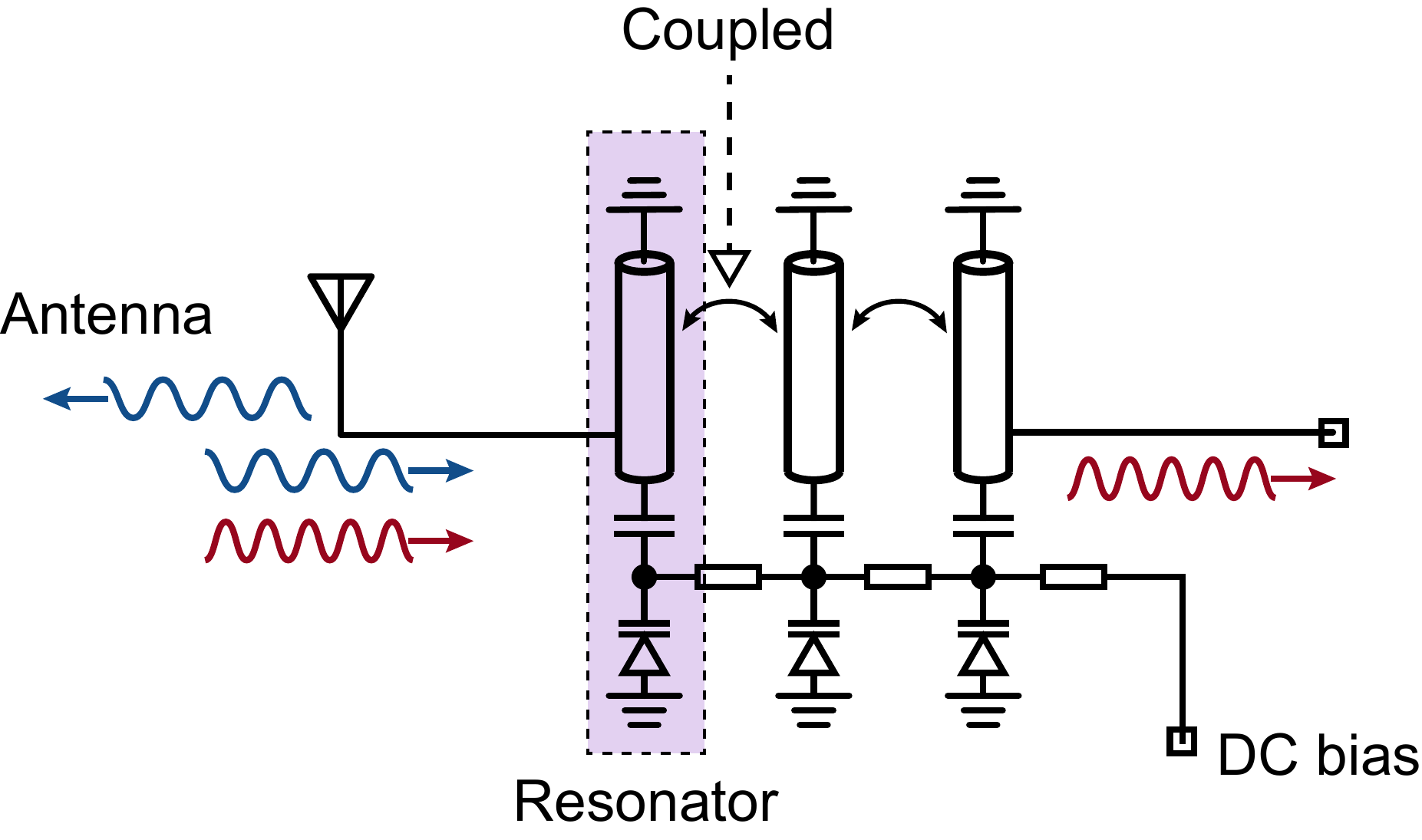}
    \caption{The structure of the frequency-tunable bandpass filter.
    }
    \label{fig:struct-bandpass-filter}
    \narrowfig
    \vspace{-4mm}
\end{figure}

As shown in Fig. \ref{fig:struct-bandpass-filter}, the bandpass filter consists of three parallel coupled tunable resonators, and each resonator has a specific resonant frequency $f$.

$f$ is determined by the two components of the resonator: a resonant bar and a resonant capacitor, according to:
\begin{equation}
    \label{equ:resonant_frequency_2}
    L=\frac{Z_0\tan(2\pi fl/v)}{2\pi f}\ \ , 
\end{equation}
\begin{equation}
    \label{equ:resonant_frequency_1}
    f=\frac{1}{2\pi\sqrt{LC}}=\frac{1}{2\pi Z_0\tan(2\pi fl/v)C}\ \ , 
\end{equation}
where $C$ represents the capacitance value of the resonant capacitor. $v$, $Z_0$ and $l$ correspond to the speed of electromagnetic waves, characteristic impedance, and length of the resonant bars, respectively.

\begin{figure}[t]
    \centering
    \subfigure[The structure of the resonator.]{
        \label{fig:reflector-resonator}
        \includegraphics[width=0.32\linewidth]{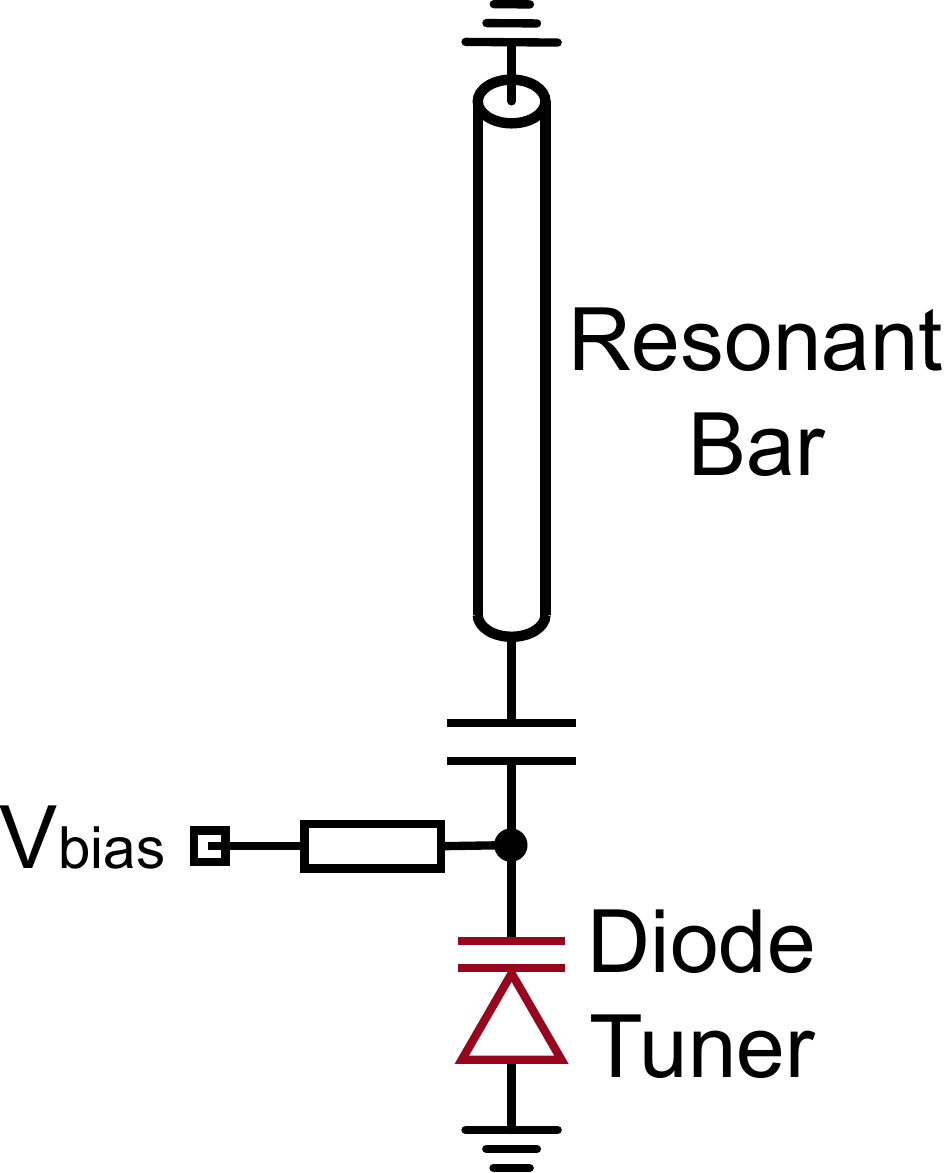}
    }    
    \hspace{0.1cm}
    \subfigure[The capacitance-voltage curve of the varactor.]{
        \label{fig:varactor-cap-vol}
        \includegraphics[width=0.58\linewidth]{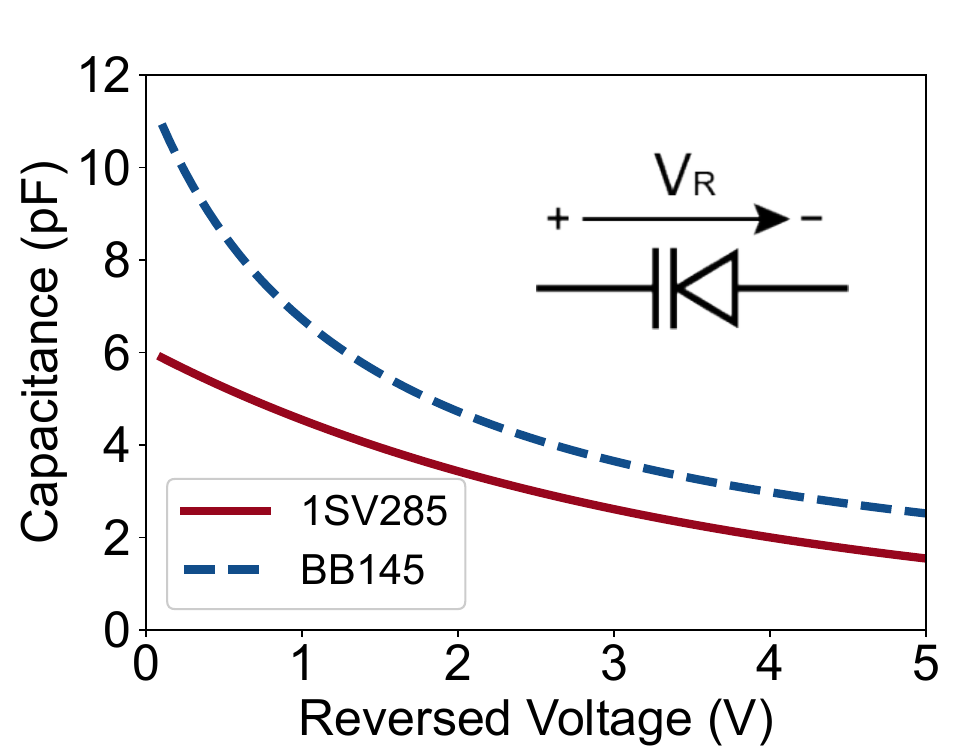}
    }
    
    \caption{\SystemName tag adjusts the $V_{bias}$ of the varactor diode to change its capacitance, thus tuning the resonator.}
    \label{fig:reflector-detail}
    \vspace{-3mm}
\end{figure}

The equation above indicates that we can change the resonant frequency by varying the resonant capacitor, thereby tuning the filter's center frequency. Thus we consider using varactors as the resonant capacitors in our tunable filter design. The varactor typically operates in a reverse-biased state, and its capacitance depends on the reversed voltage, according to:
\begin{equation}
    C=\frac{C_0}{(1-V/V_0)^{\gamma}}
\end{equation}
where $C_0$ is the junction capacitance with no bias. $V_0$ and $\gamma$ depend on the diode type and are constants for a specific diode. Specifically in Fig. \ref{fig:varactor-cap-vol}, we show the junction capacitance versus the voltage for diode 1SV285 \cite{datasheet1sv285} and BB145 \cite{datasheetbb145}. By using the varactor, we can adjust the resonant capacitance value by providing a set of different bias voltages, further tuning the center frequency of the bandpass filter among several bands. Furthermore, the varactors only consume a current of several nAs, introducing negligible additional power consumption to the tag.

To provide different bias voltages to the tunable bandpass filter, a resistor voltage divider circuit is employed. This circuit utilizes an SPNT (i.e., single-pole N-throw) analog switch to connect various resistances to the voltage divider circuit. By manipulating the analog switch, the voltage divider circuit generates different bias voltages, thereby facilitating the adjustment of the filter's center frequency.

The frequency-tunable bandpass filter is employed in the next stage of the antenna. When the frequency of the input signal deviates significantly from the resonant frequency, the resonators fail to resonate, resulting in weak electromagnetic oscillation. Consequently, the input excitation signal is unable to traverse to the terminal through the coupling and instead gets reflected back by the filter.

When the frequency of the input RF signal matches the resonant frequency, the amplitude of the electromagnetic field oscillation around the resonator is significant. This allows the signal to be coupled and transmitted to the adjacent resonator, even if the two resonators are not directly connected. Through successive coupling, the signal propagates and eventually passes through the entire filter. Thus the signal in a specific frequency band is extracted.

\water\subsubsection{Two-step Strength Comparison}
After the signal extraction, we use an envelope detector to convert the strength of the excitation signal at that band into a voltage output. 
By comparing the output voltages across the different frequency bands, the tag can identify the band with the strongest excitation signals. 

Instead of employing ADC to sample the output voltages and comparing them with each other, which would augment tag complexity and lead to increased energy consumption, we propose a method in which the output voltages are stored in capacitors before being compared. This approach avoids the need for directly obtaining accurate voltage values, and by using pairwise comparisons, the tag can identify the frequency band with the strongest excitation signal.

\begin{figure}[!t]
    \centering
    \subfigure[\textit{Step1:} Charge.]{
        \label{fig:comparator-charge}
        \includegraphics[width=0.96\linewidth]{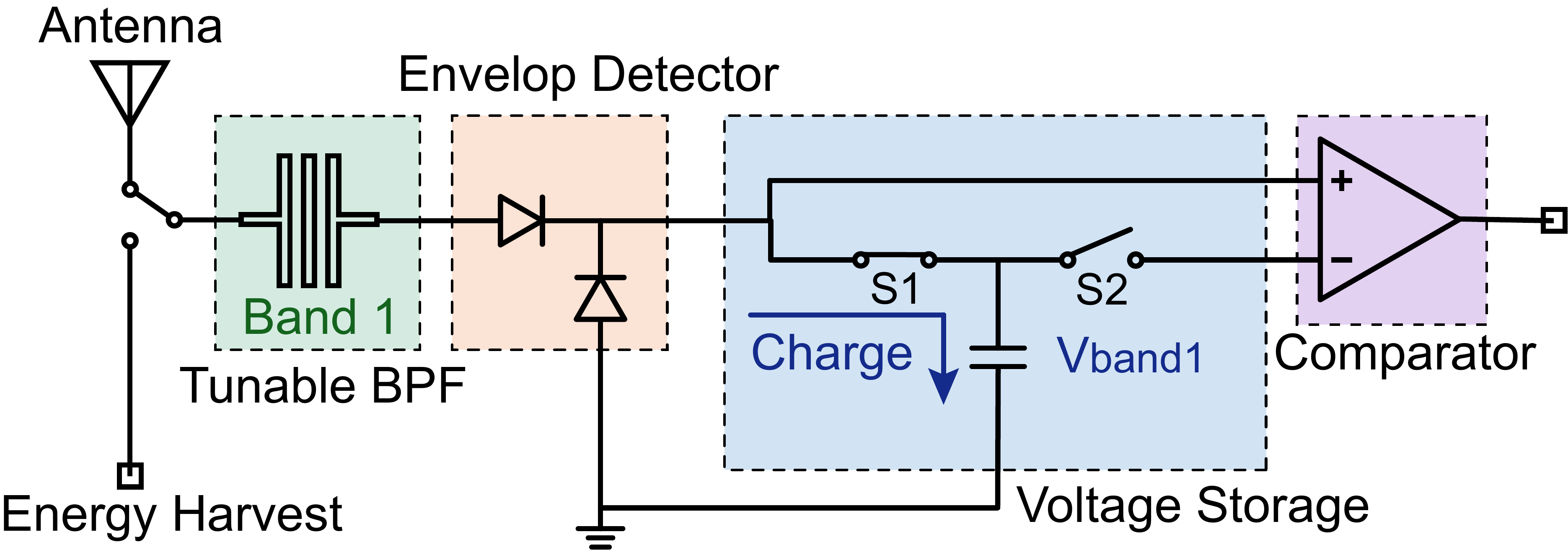}
    }
    \subfigure[\textit{Step2:} Compare.]{
        \includegraphics[width=0.96\linewidth]{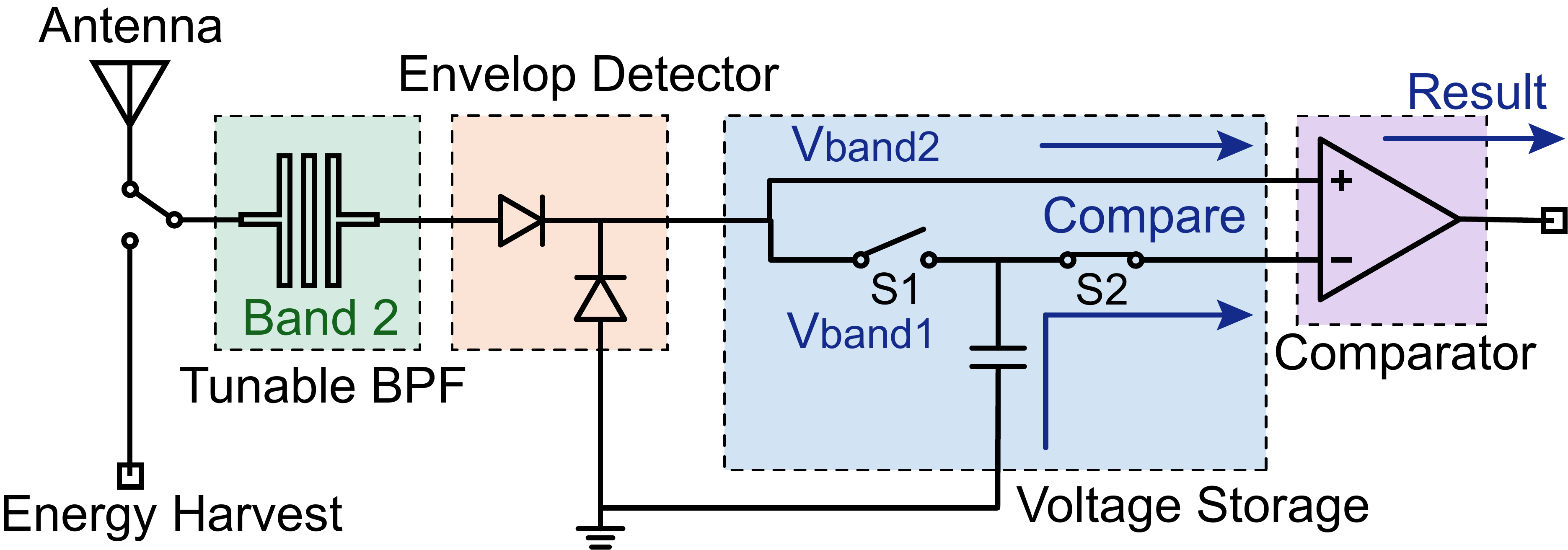}
        \label{fig:comparator-compare}
    }
    \caption{The frequency band detector completes the comparison of signal strengths at two frequency bands through two steps.}
    \label{fig:comparator-work-flow}
    \narrowfig
    \vspace{-5mm}
\end{figure}

Our frequency band detector performs a comparison of the excitation signal strength in two frequency bands in two steps, as illustrated in Fig. \ref{fig:comparator-work-flow}. Firstly, the center frequency of the tunable bandpass filter is adjusted to \textit{Band 1}, and then switch S1 is closed while switch S2 is open. At this stage, the envelope detector converts the excitation signal strength in \textit{Band 1} into voltage $V_{band1}$, and the capacitor is charged to $V_{band1}$. Next, the center frequency of the tunable bandpass filter is adjusted to \textit{Band 2}, and S1 is opened while S2 is closed. Now the inputs of the comparator are the voltage $V_{band1}$ stored in the capacitor, representing the signal strength of \textit{Band 1}, and the voltage $V_{band2}$ outputted by the envelope detector, representing the signal strength of \textit{Band 2}. The output level of the comparator indicates the result of the comparison between the signal strength of the two bands: if the output level is high, it means \textit{Band 2} has a stronger signal, otherwise, \textit{Band 1} has a stronger signal. By performing two comparisons, this circuit can determine the frequency band with the strongest excitation signal among the three frequency bands.

Besides, the capacitor could experience slight voltage reduction due to the leakage current of the comparator and analog switches. This leads to potential inaccuracies in the voltage comparison, especially when the two voltages are very close. Thus, The tag will exchange the order of two voltage inputs and then perform another comparison. If the result of the two comparisons differs, the tag will consider the signal strengths in those two frequency bands to be practically the same and randomly choose one as the comparison result.

\subsection{Frequency-selective Reflector}
\label{title:frequency-selective-reflector}


We consider the frequency-selective reflection as the capability of generating a strong backscattered signal only at a specific frequency band while producing weak signals in other frequency bands. For the methodology of generating reflection signals of different strengths in different frequency bands, we first analyze the mathematical model of the backscattered signal and study the factors that influence the strength of the backscattered signal.


When the tag controls the RF switch to toggle between two terminal loads. The signal received by the reader can be represented as:
\begin{equation}
    y(t)=\alpha x(t)+\beta B(t)x(t)+n(t)
\end{equation}
where $x(t)$ is the reader's excitation signal, $n(t)$ is the noise, $\alpha$ and $\beta$ are the complex attenuation of the excitation signal and the backscattered signal, and the $B(t)$ are either $\Gamma_1$ or $\Gamma_2$, which are complex reflection coefficients corresponding to the two terminal loads.
%
The average of the second term, $\beta(\Gamma_1+\Gamma_2)x(t)/2$, corresponds to the unchanged reflection of the excitation signal and does not carry any tag information. Thus when calculating the strength of the backscattered signal, this part should be subtracted, and the strength should be written as $|\beta(\Gamma_1-\Gamma_2)x(t)/2|$. 
This implies that the strength of the backscattered signal depends on the difference in reflection coefficient when the tag switches its terminal loads. 
Therefore, if we ensure that the reflector exhibits a significant variation of reflection coefficient only at one frequency band, we can achieve the frequency-selective reflection.

According to the characteristic of the tunable bandpass filter described in the previous part, when the input signal frequency matches the center frequency of the filter, the signal can reach the switchable terminal loads, thus the reflection coefficient of the entire reflector for that frequency signal is primarily determined by the impedance of the terminal loads. For signals with frequencies different from the filter's center frequency, they are directly reflected back by the filter, thus the reflection coefficient of the reflector for that frequency signal is determined by the filter and remains unchanged when terminal loads are switched.

\begin{figure}[tb]
    \centering
    \includegraphics[width=0.7\linewidth]{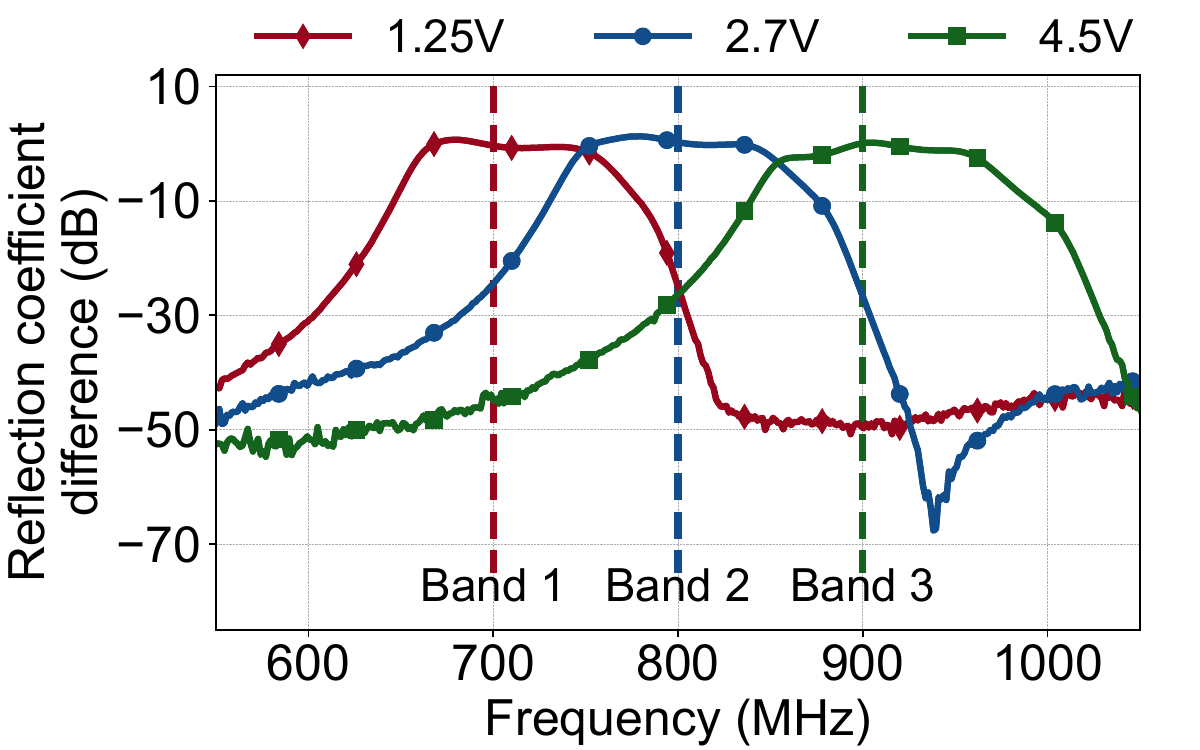}
    \caption{The operating frequency band of the reflector changes with the variation of the bias voltage.}
    \label{fig:reflector-test}
    \narrowfig
    \vspace{-3mm}
\end{figure}

Based on the above analysis, we consider adding a voltage-tunable bandpass filter mentioned above in front of the traditional RF switch to form a frequency-selective reflector. To confirm the feasibility of this approach, we measure the difference in reflection coefficients in various frequency bands when the reflector toggles the terminal loads. Fig. \ref{fig:reflector-test} illustrates the difference in reflection coefficients for two different terminal loads corresponding to three different bias voltage inputs applied to the reflector. Although the reflection coefficient difference of the reflector under three bias voltage inputs may not be completely isolated from each other, each peak corresponds to only one frequency band, and the changes in reflection coefficient for the other two bands are minimal. This indicates that when the reflector selects a specific frequency band, it will generate a strong backscattered signal only at that band, while the backscattered signal in the other frequency bands is extremely weak. Hence, this reflector exhibits excellent capability of frequency-selective reflection.

\subsection{Reflection Power Adjuster}





In backscatter communication, the strength of backscattered signals is directly related to the strength of the excitation signal, which varies with the distance between the tag and the reader. In the case of a tag being close to a reader, its backscattered signal can be excessively strong and cause interference to distant readers operating at the same frequency. To control the strength of backscattered signals, the tag first needs to detect the strength of the excitation signal. Thus, we introduce the excessive power detector.

\begin{figure}[tb]
    \centering
    \includegraphics[width=0.96\linewidth]{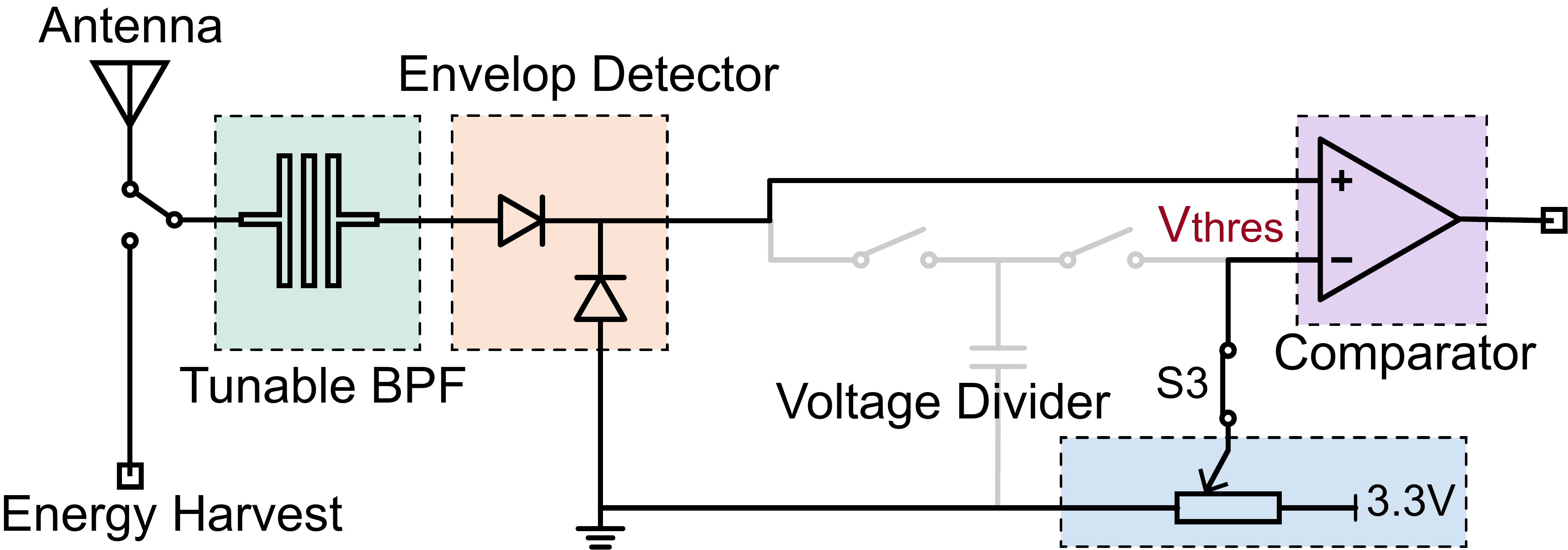}
    \caption{The excessive power detection circuit design based on the frequency band detector.}
    \label{fig:high-power-detector}
    \narrowfig
    \vspace{-3mm}
\end{figure}

Noticing that the envelope detector converts the strength of the excitation signal into a voltage output, allowing for checking whether the excitation signal is excessively intense through a threshold voltage comparison. Considering the frequency band detector already includes an envelope detector and voltage comparator, we add a threshold voltage comparison circuit to the frequency band selector to implement the excessive power detector, as shown in Fig. \ref{fig:high-power-detector}.
This design can also help to simplify the tag design and reduce power consumption.

When S3 is closed, the threshold voltage comparison circuit is connected and the detector enters excessive power detection mode. At this point, the comparator's inputs are the threshold voltage from the voltage divider circuit and the detection voltage from the envelope detector. 
\rev{Using the link budget theory\cite{griffin2009complete}, we calculate that when the excitation signal is stronger than -20dBm, the tag can interfere with the adjacent readers. Thus, the threshold voltage is set to 1.6V, which is the output voltage of the envelope detector when the excitation is -20dBm\cite{datasheetlt5534}.} Finally, by comparing the detected voltage with the threshold voltage, the tag determines whether the current excitation signal is too strong.

\begin{figure}[tbp]
    \centering
    \subfigure[The schematic of the multiple selectable impedances.]{
        \includegraphics[width=0.55\linewidth]{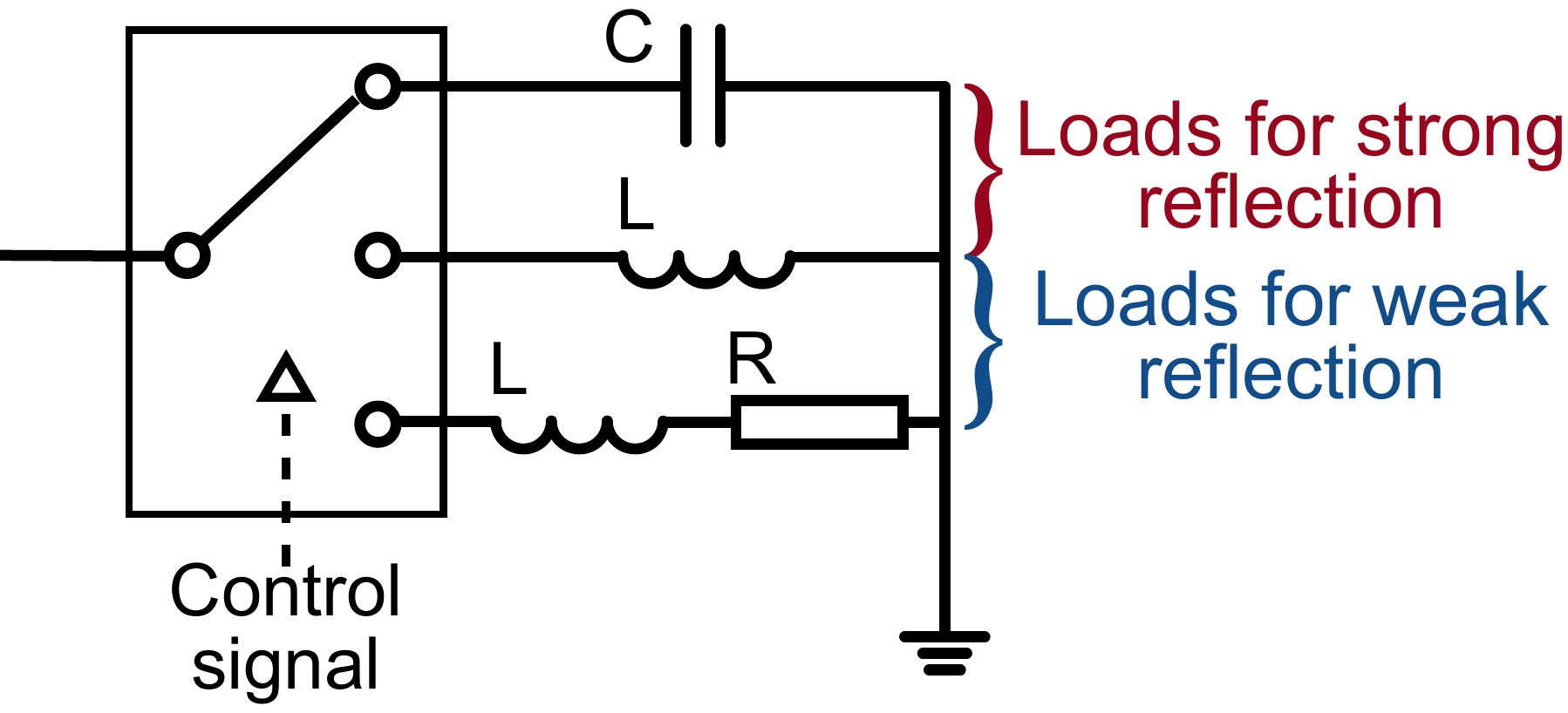}
        \label{fig:schematic-multi-select-loads}
    }
    \hspace{0.1cm}
    \subfigure[The reflection coefficients of the impedances of L, C and series of L and R.]{
        \includegraphics[width=0.3\linewidth]{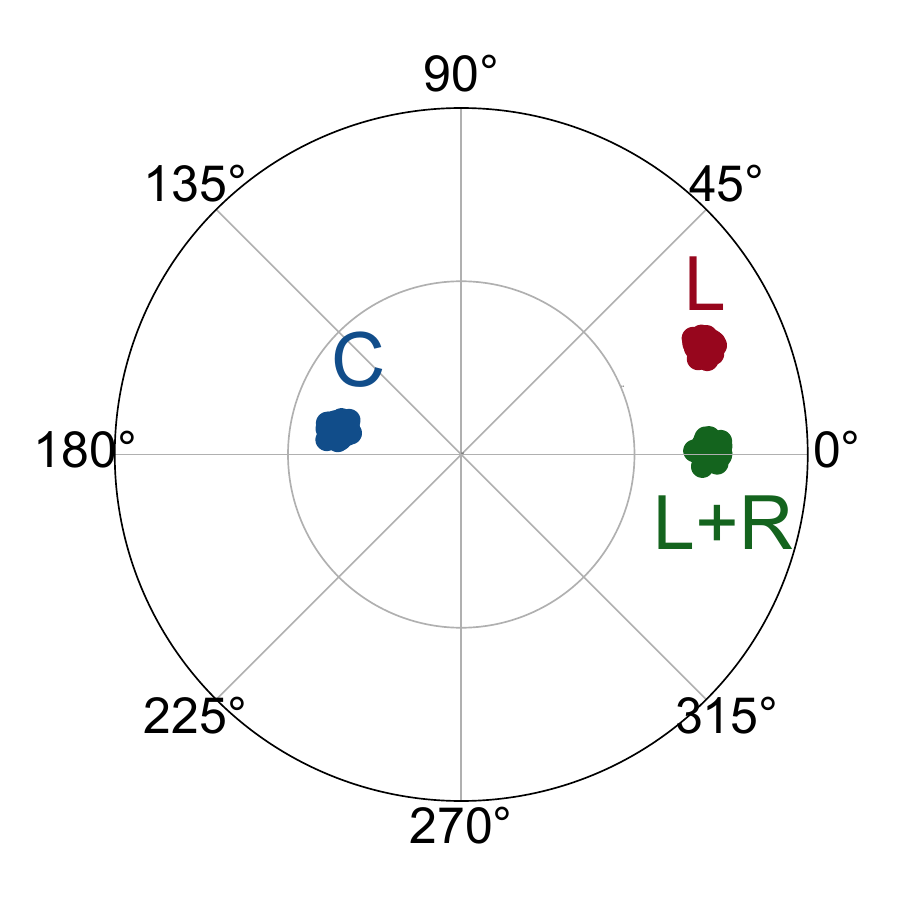}
        \label{fig:smith-multi-select-loads}
    }
    \caption{The multiple impedance selection circuit and the reflection coefficients for each impedance.}
    \label{fig:multi-terminal}
    \narrowfig
    \vspace{-3mm}
\end{figure}

Once the excessive power detector detects the strength of the excitation signal, the tag needs to adaptively adjust the strength of the backscattered signal based on the detection result. According to the expression for the backscattered signal strength described in the previous subsection, we can adjust the strength of the reflected signal by varying the difference in the reflection coefficients of the reflector. 

Within the selected frequency band of the reflector, the reflection coefficient $\Gamma$ for the input signal primarily depends on the impedance of terminal loads, according to the following equation:
\begin{equation}
    \label{equ:gamma}
    \Gamma=\frac{Z_0-Z_L}{Z_0+Z_L}
\end{equation}
where $Z_0$ is the characteristic impedance which is commonly 50$\Omega$, and $Z_L$ is the complex impedance of the terminal load. \add{This equation can also be used to calculate the terminal impedance given the required reflection coefficients.} We employ multiple selectable terminal loads with proper impedance on the tag's reflector. 
\rev{We carefully adjust the value of the load impedance to maximize the differential reflection coefficient for the higher SNR.} The loads we designed include an inductor, a capacitor, and a series of an inductor and a resistor, as shown in Fig. \ref{fig:multi-terminal}. By selecting appropriate loads based on the strength of the excitation signal, the backscattered signal's strength can be controlled effectively.

When the strength of the excitation signals does not exceed the threshold, the terminal loads are switched between the inductor and the capacitor. Due to the opposite imaginary part of the impedance of the inductor and the capacitor, the variation in the reflection coefficient is relatively significant when switching between these two loads. It helps strengthen the backscattered signal, improving the communication performance between the tag and the reader.


When the excessive power detector detects an excessively strong excitation signal, the tag's loads are switched between the inductor and the series of an inductor and resistor. Because of the small difference in impedance of these two terminal loads, the variation in the reflection coefficient during this switch is relatively smaller, thereby reducing the strength of the reflected signal. This helps avoid interference with distant readers operating in the same frequency band.

\addsec
\section{Reader Frequency Assignment}
\label{title:reader-frequency-assignment}

\SystemName is required to operate in a backscatter network where the adjacent readers work on different bands. In this section, we propose an algorithm to assign the readers' working frequency bands, separating the same-band readers to avoid interference. 


In an ideal scenario, we can employ a hexagonal grid deployment similar to LTE base stations\cite{novlan2011analytical}. 
However, the deployment of readers is often constrained by various factors in real IoT environments. We turn to carefully assign the operation frequencies to readers for interference avoidance.


We model the aforementioned issue as a coloring problem in an undirected graph. We construct an interference graph. Each node in the graph represents a reader. If the interference strength between two readers exceeds a safety threshold, an undirected edge is established between the nodes representing these two readers. We aim to color the nodes using three colors, intending for adjacent nodes to have different colors. This requires the graph to be a tripartite graph. However, due to the uncertainty in reader positions and the wireless channel, we can't guarantee that the graph is tripartite. Consequently, there might not exist an allocation method that entirely avoids interference.


We shift our focus towards finding the assignment with the least co-channel interference, which is an NP-hard problem. Traversing through all possible solutions leads to a complexity of $O(3^n)$, which is impractical. Therefore, we propose an optimization algorithm, as shown in Alg. \ref{alg:frequency-assign}. This algorithm takes the interference strength between readers as its input, with the frequency assigned to each reader as the optimization object, aiming to minimize the strength of co-frequency interference. Considering that frequency allocation is an optimization problem with discrete variables, we employed a genetic algorithm. The algorithm unfolds through the following steps:

\begin{algorithm}[t]
    \SetKwRepeat{Do}{do}{while}
    \SetAlgoLined
    \caption{Reader\ Frequency\ Assignment\ with\ Genetic\ Algorithm}
    \label{alg:frequency-assign}
    \KwIn{$M_{ij}$: Interference strength between readers}
    \KwOut{$F$: Reader frequency assignment strategy}
    \BlankLine
    Initialize interference graph's adjacency matrix: $G_{ij}=(M_{ij}>Thres)?M_{ij}:0$\;
    Initialize population with random individuals (assignments): $P_0=[F_0, F_1, \cdots, F_{M-1}]$\;
    $cnt \gets 0$\;
    \Do{$cnt < C$}{
        \For{$F_{i} \text{ in } P_{cnt}$}{
            \ForEach{Reader $R_j$}{
                Find same-band readers' index $K$\;
                $E_j \gets \sum_{k\in K}G_{jk}$\;
                $Error_i \gets \text{mean}(E_j) + \text{median}(E_j)$\;
            }
        }
        $P_{cnt} \gets \text{top 10\% individuals of } P_{cnt}$,
        ascending sorted by $Error_i$\;
        $P_{cnt+1} \gets \text{mate, mutate, expansion }(P_{cnt})$\;
        $cnt \gets cnt + 1$\;
    }
    $F \gets F_{i} \text{ in } P_C \text{ with minimum } Error_i$\;
\end{algorithm}

\begin{enumerate}
    \item First, we initialize the edge weight of graph $G_{ij}$ as the interference strength between the corresponding readers at both ends (Line 1). Subsequently, we initialize the frequency assignment $F_i$ as individuals in population $P_0$, where $i \in [0, M-1]$ (Line2).

    \item For each assignment method $F_{i}$ within the population, we compute the total interference energy received by each reader, represented as $E_{j}$, i.e., the node strength (Line 7-8). We then utilize the sum of the maximum value and the median of $E_{j}$ as the error. This selection considers both local and global optimization simultaneously (Line 9).

    \item We select the top 10\% individuals with the lowest error and use mating, mutation, and population expansion methods to generate the next generation $P_{cnt+1}$. We proceed with the next round of evaluation until the iteration count reaches the preset limit (Line 12-14).

    \item Finally we output the individual with minimum error in population $P_{C}$ as the optimized reader frequency assignment strategy (Line 16).
\end{enumerate}





This algorithm requires the interference strength between readers to construct the interference graph. We set a bootstrap phase, during which one reader sequentially transmits test signals and other readers listen to the channel and record interference strength. \rev{This phase takes an average of 2.3s when the algorithm runs in an AMD R7-4800H CPU and will not impact the subsequent readings.}

\rev{To adapt to the dynamic channel, we introduce a reallocation mechanism. After the frequency allocation, each reader listens and detects the interference from other readers, which is denoted as $\vec{v_0}$. And they periodically detect and report the current interference strength, denoted as $\vec{v_i}$. When the normalized dot product of $\vec{v_0}$ and $\vec{v_i}$ is less than the threshold, we consider that the channel has changed and reallocate the frequencies.}

\section{Practical Issue}
\label{title:practical-issue}

We complete the design of \SystemName by articulating some practical issues of its applicability.

\subsection{Bandpass Filter Design}


We incorporate two frequency-tunable bandpass filters in \SystemName tag, each serving distinct purposes: one for extracting the excitation signal and the other for generating backscattered signals within specific frequency bands. However, when the Q value of the bandpass filter remains constant, increasing the out-of-band suppression inevitably leads to greater in-band losses. Higher losses within the passband significantly weaken tag communication performance, while lower out-of-band suppression increases the risk of interfering with other frequency bands.

We set an out-of-band suppression target of 20dB. Leveraging the ANSYS HFSS optimization, we aimed to minimize in-band losses while ensuring out-of-band suppression. Additionally, we tailored the optimization objectives for the frequency band detector and frequency-selective reflector to focus on the $S_{21}$ and $S_{11}$ parameters of the filters, respectively, catering to the distinct purposes of the two bandpass filters.


\subsection{Envelope Detector Selection}

In prior work, researchers utilize passive \cite{guo2022saiyan, li2022passive} or active \cite{zhang2016hitchhike} envelope detectors for the tags. However, the active envelope detector often consumes several mWs, exceeding the power constraints of the backscatter tag. While the passive envelope detector can output detection voltage without additional power consumption, it suffers from lower sensitivity. As the frequency band detection step in \SystemName tag requires precise detection voltage output, the passive envelope detector is not suitable for our design.


Reviewing the band detector's design, we noticed that the envelope detector works for only tens of microseconds to complete the band detection, instead of requiring continuous operation as seen in ambient backscatter tags \cite{chi2020ltebackscatter}. By setting the band detection frequency to once every 3 seconds, the envelope detector operates at less than a 1\textperthousand~duty cycle, resulting in an average power consumption of just a few $\mu$Ws.

\subsection{Capacitor Selection}


In the frequency band detector, we utilize a capacitor to store the detection voltage. If the capacitance value is too small, leaked currents from switches and comparators can rapidly reduce its voltage, leading to erroneous signal strength comparison results. On the other hand, if the capacitance value is too large, the active envelope detector would need to operate for an extended period until the capacitor's voltage stabilizes, incurring higher power consumption.

We aim for the error between the capacitor voltage and the detector output voltage to be within 1\%, after charging the capacitor for $T_c$ seconds and considering leakage for $T_l$ seconds. We can compute the upper and lower bounds of the capacitance value using classical circuit theory.:
\begin{equation}
    C \leq C_{max} = \frac{T_c}{\ln{(100)}R_S} \quad
    C \geq C_{min} = \frac{100I_lT_l}{V_{min}} \,
\end{equation}
where $R_s$, $I_l$, and $V_{min}$ are the output impedance of the envelope detector, the leak current of the CMOS switch, and the minimum valid output voltage of the envelope detector, respectively. \rev{We set $T_c$ to 50 $\mu$s and adopt a capacitor of 10nF, which is the geometric mean of the capacitance values $C_{max}$ and $C_{min}$. If we need to further reduce the energy consumption of the envelope detector, we can set a lower $T_c$ and use a smaller capacitor.}

\color{black}
\section{Implementation}
\label{title:implementation}

\begin{figure}[t]
    \centering
    \includegraphics[width=0.75\linewidth]{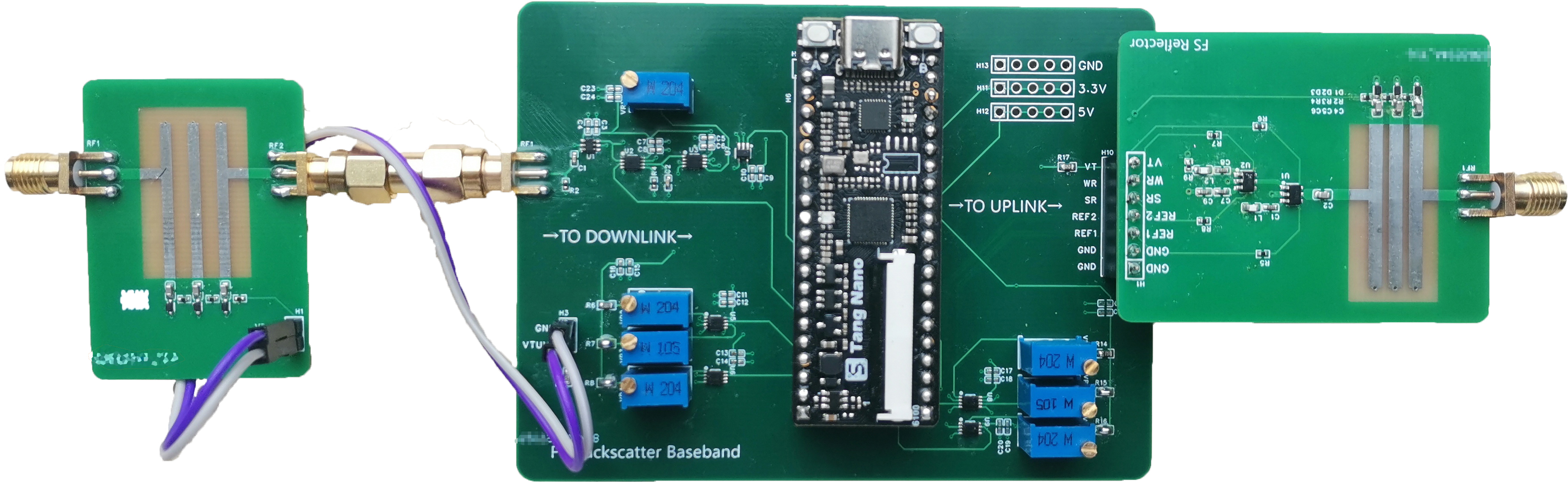}
    \caption{\SystemName tag is implemented by commercial off-the-shelf components on PCB. }
    \label{fig:backscatter_tag_pcb}
    \narrowfig
    \vspace{-4mm}
\end{figure}

\subsection{\SystemName Tag}

We implement the \SystemName tag on PCB using commercial off-the-shelf components, as shown in Fig. \ref{fig:backscatter_tag_pcb}. The tag consists of a frequency band detector with excessive power detection and a frequency-selective reflector with adjustable reflection strength.

In the frequency band detector circuit, the first component is the frequency-tunable bandpass filter. The filter is implemented using microstrip lines integrated on the PCB and three 1SV285 varactor diodes. To provide an adjustable bias voltage to adjust the center frequency of the filter, we use analog switches TS5A23166 to switch the resistors in the voltage divider circuit. For the two-step strength comparison circuit, we use the LT5534 envelope detector, \rev{which can effectively detect RF power above -40dBm.} We control LT5534's EN pin to enable this chip only during the frequency band detection period for energy saving. The TS5A23166 analog switch is employed to toggle among the charging and two comparing modes (comparing with capacitor voltage or threshold voltage), and an NCS2200 low-power comparator is utilized for voltage comparison. This circuit enables the comparison of excitation signal strengths in two frequency bands and detects excessively strong excitation signals.


The reflector is composed of the tunable bandpass filter, the switchable terminal, and the control logic. The tunable bandpass filter is similar to that in the reflector. The switchable terminal utilizes two HMC544 RF switches to toggle among three terminal impedances, which can adjust the strength of the backscattered signals. \rev{Both the detector and the reflector are equipped with 3-dBi omnidirectional antennas.}

The control logic is implemented using a low-power GW1NZ-1 FPGA. The FPGA utilizes square waves of different phases at a frequency of 1MHz to switch the terminal loads. This process generates a BPSK backscattered signal at the frequency band adjacent to the excitation signal at 1MHz to avoid self-interference. 

\rev{The cost of making a PCB prototype is around \$13. Compared to the existing backscatter tag designs, \SystemName's additional costs come from the varactor diodes in the tunable filter and the analog switch, which cost less than \$1. Thus \SystemName is not expensive compared to the existing designs.}




\vspace{-2mm}
\subsection{Reader}

We implement the mono-static configured readers using the USRP N210 software-defined radio platform by Ettus Research. \rev{The USRPs are equipped with two 3-dBi co-located antennas separated by 30cm. The transmission antenna is connected to a UBX-40 daughterboard. To prevent excessive self-interference from damaging the RX chain, the reader transmission power is limited to +5 dBm.} The receiving antenna is connected to the same daughterboard, which down-converts the received signals to the baseband and samples at a rate of 500 ksps. \rev{The reader can read tags within 1m. If we can cancel the self-interference in the RX chain like the commercial RFID readers, we could increase the transmission power and achieve a longer communication range.}

\vspace{-2mm}
\subsection{Power Consumption}

\begin{table}[t]
    \centering
    \renewcommand{\arraystretch}{1.1}
    \caption{Power consumption of \SystemName tag implemented in 65nm CMOS technology and the PCB prototype}
    \label{tab:energy-consumption}
    \begin{tabular}{ccccc}
        \hline
        Part & Switches & Envelope detector & \multicolumn{2}{c}{Control logic} \\
        \hline
        ASIC Power & 3$\mu$W & 7$\mu$W & \multicolumn{2}{c}{2$\mu$W}\\
        PCB Power & 3$\mu$W & 0.25mW & \multicolumn{2}{c}{16.6mW}\\
        \hline
        \noalign{\vskip 1mm}
        \hline
        Part & Comparator & Filter tuning & Clock & Total\\
        \hline
        ASIC Power & 1$\mu$W & 60nW & 3$\mu$W & 16$\mu$W\\
        PCB Power & 1.6mW & 60nW & 12mW & 30.45mW\\
        \hline
    \end{tabular}
    \narrowfig
\end{table}

\begin{table}[t]
    \centering
    \renewcommand{\arraystretch}{1.1}
    \caption{Power Consumption and Energy Efficiency Comparison of Trident and Other Backscatter Technologies}
    \label{tab:energy-comparison}
    \begin{tabular}{ccc}
        \hline
         Technology & Power consumption& Energy efficiency\\
         \hline
         OFDMA Backscatter \cite{zhao2019ofdma} & 18.31$\mu$W & 0.079nJ/bit\\
         NetScatter \cite{hessar2019netscatter} & 45.2$\mu$W & 45.2nJ/bit\\
         P2LoRa \cite{jiang2021long} & 320 $\mu$W & 28.39nJ/bit\\
         mmComb \cite{yoon2024mmcomb} & 87.3$\mu$W & 1.59pJ/bit\\
         \hline
         \textbf{Trident} & 16$\mu$W & 0.16nJ/bit \\
         \hline
    \end{tabular}
    \vspace{-5mm}
\end{table}

\rev{We evaluate the power consumption of an ASIC solution of the \SystemName tag \cite{na2023leggiero, dunna2021syncscatter} and the PCB prototype. We report it in Tab.\ref{tab:energy-consumption}. \SystemName with ASIC technology meets the energy constraints of backscatter tags with the power consumption of 16$\mu$W. The PCB prototype implementation has a power consumption of about 30mW. We also compare the power consumption and the energy efficiency of \SystemName and other backscatter technologies, which are enabled to avoid interference. We report it in Tab.\ref{tab:energy-comparison}.} 

\section{Evaluation}
\label{title:evaluation}

\begin{figure}[t]
    \centering
    \includegraphics[width=0.75\linewidth]{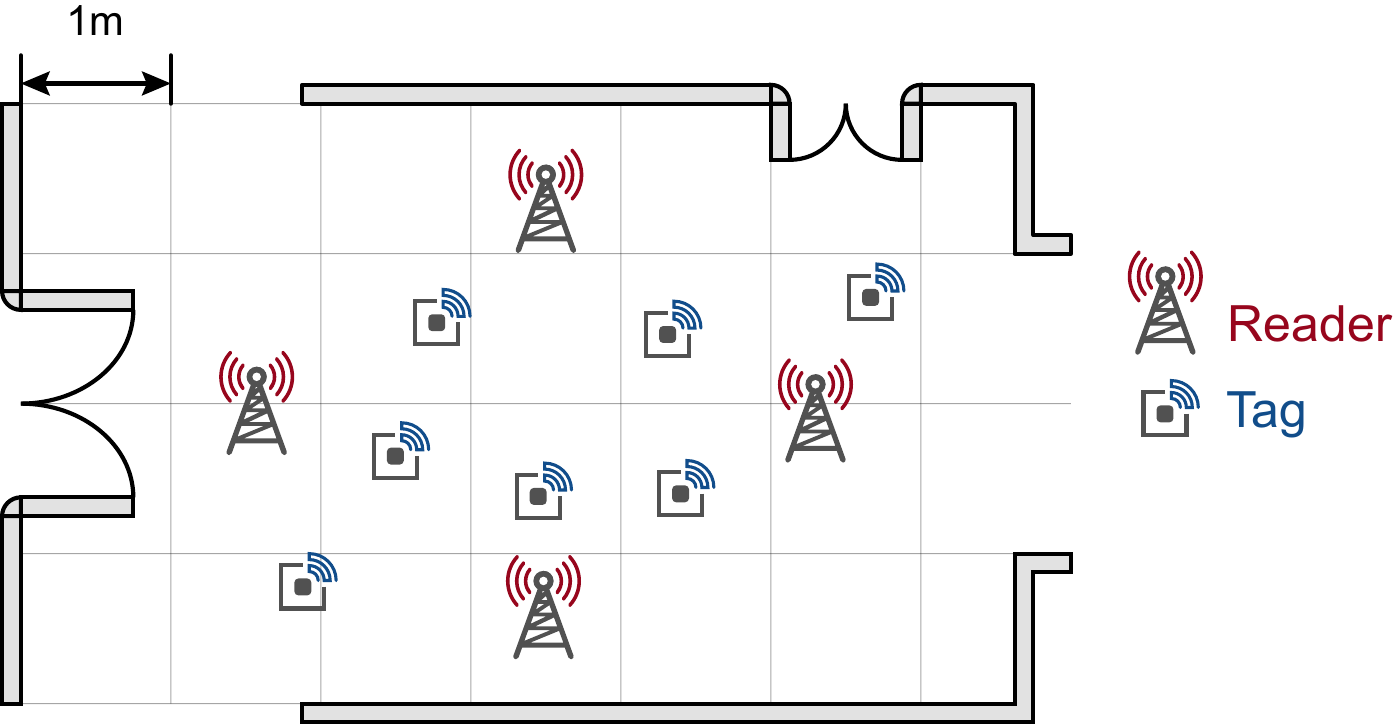}
    \caption{Evaluation setting in a 4m$\times$6m corridor.}
    \label{fig:evaluation-setting}
    \narrowfig
    \vspace{-3mm}
\end{figure}

\begin{figure*}[t]
    \centering
    \subfigure[Deploy 4 tags]{
        \includegraphics[width=0.23\linewidth]{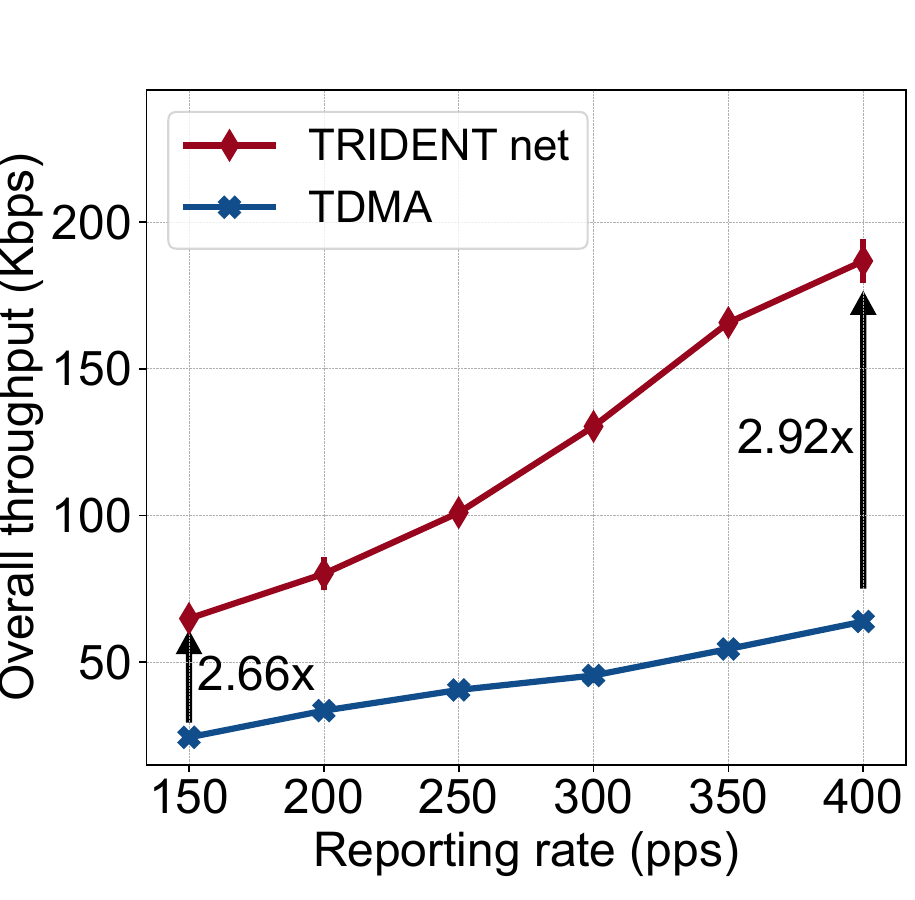}
    }
    \hspace{-0.2cm}
    \subfigure[Deploy 5 tags]{
        \includegraphics[width=0.23\linewidth]{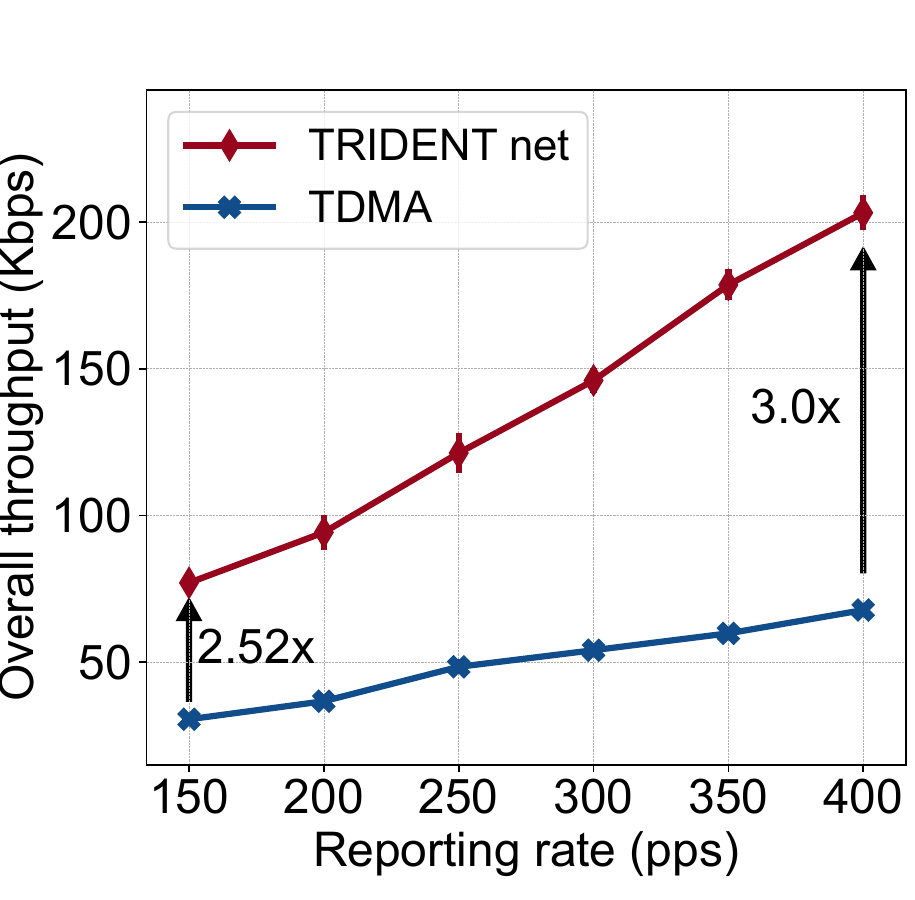}
    }
    \hspace{-0.2cm}
    \subfigure[Deploy 6 tags]{
        \includegraphics[width=0.23\linewidth]{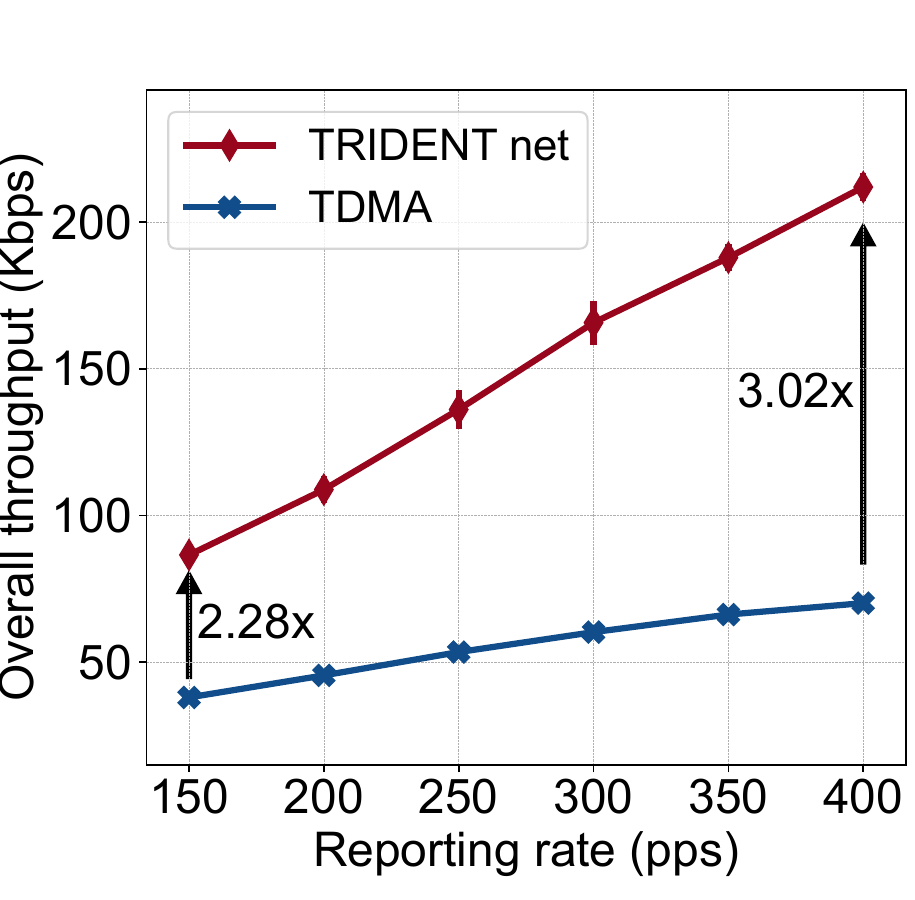}
    }
    \hspace{-0.2cm}
    \subfigure[Deploy 7 tags]{
        \includegraphics[width=0.23\linewidth]{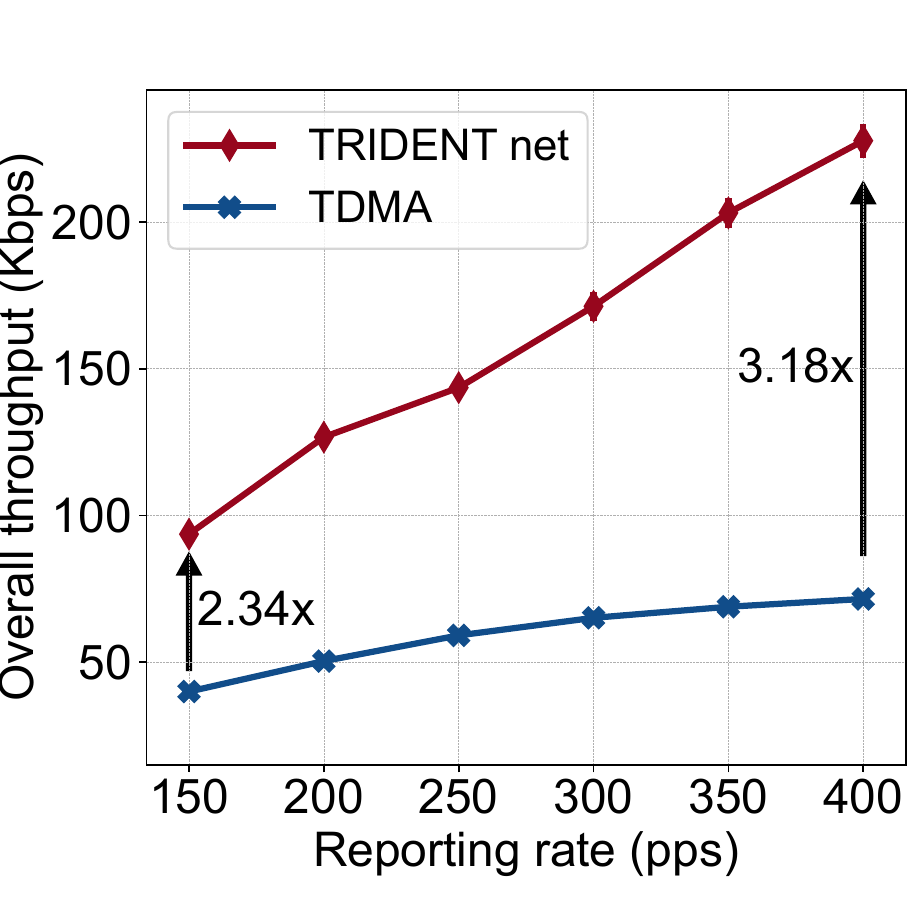}
    }
    \caption{Overall throughput of \SystemName network and traditional TDMA network in different number of tags deployed in the network and reporting rate of each tag.}
    \label{fig:N_tag_throughput_vs_reporting_rate}
    \vspace{-9mm}
\end{figure*}

\subsection{Experimental Setup}

We construct a \SystemName network prototype with 4 readers and 7 tags in an indoor corridor measuring 4m$\times$6m. As shown in Fig. \ref{fig:evaluation-setting}, 
the 4 readers were deployed around the center of the corridor and the 7 tags were randomly placed within the coverage area. 
The upper and lower readers in the figure operate at the same band, while the remaining two readers each use the remaining two bands, respectively.
And when these tags detect the excitation signal, they would transmit a data packet of 128 bits at a rate of 100Kbps with a variable reporting rate from 100 packets per second(pps) to 500pps.

We present the overall evaluation results first. In the overall system evaluation, we take \textit{overall throughput} as the key metric to assess the \SystemName network. \textit{overall throughput} measures the average amount of backscattered data correctly decoded per second at all the readers in the network. Then we present our ablation study, where we evaluate the frequency selectivity, the accuracy of band selection, and the throughput increment by adjusting the reflection power. \add{We also conduct simulation experiments to evaluate the frequency allocation algorithm.}

\subsection{Overall Throughput}
\naxin {We evaluate the overall throughput of the \SystemName network in terms of different tag numbers, reporting rates, reader density, and tag deployment. We construct a centralized TDMA network as the baseline and show \SystemName's performance gain. The CSMA network is not compared here since its media access efficiency is typically lower than TDMA\cite{sharp1995hybrid}.} \rev{Both networks use the same deployment plan of all the readers and tags to ensure that they are the same in terms of the SNR level.}

\subsubsection{Impact of Tag Numbers And Reporting Rate}

We conduct four groups of experiments by varying the number of tags from 4 to 7. For each tag number setting, we vary the tags' reporting rate from 150pps to 400pps and measure the overall throughput. We have three observations based on the results shown in Fig.\ref{fig:N_tag_throughput_vs_reporting_rate}.

First, we observe that \SystemName demonstrates a throughput improvement of 2.28-3.18$\times$ across various numbers of tags and reporting rate settings compared to the TDMA network. This is because readers do not need to operate intermittently in \SystemName, resulting in a three times increment in their operating time. 



Second, we observe that the total throughput of \SystemName network increases linearly with the tags' reporting rate, while the throughput increase of the TDMA network gradually slows down. 
For instance, when the tag number is 6, for every 50pps increase in the reporting rate, the \SystemName network's throughput increases linearly by nearly 25Kbps. However, the throughput increment of the TDMA network decreases from 7.5Kbps to 3Kbps. 
The performance gain also increases with the reporting rate, ranging from 2.28-2.66$\times$ at a reporting rate of 150pps to 2.92-3.18$\times$ at a reporting rate of 400pps.

Third, when the reporting rate is high, the total throughput of \SystemName network continues to grow with the increasing tag number, while the increase in total throughput of the TDMA network is highly limited. At a reporting rate of 400pps, the average increase in throughput of the \SystemName network is 13.9Kbps per additional tag. (27.8\% of the ideal throughput per tag). On the other hand, TDMA network only achieves an average increase of 7.7Kbps in throughput for each additional tag (15.4\% of the ideal throughput per tag).

The reason is that a reader in the TDMA network can excite the tags which wouldn't respond to it in \SystemName network. This results in the reader communicating with more tags, and these tags cause more serious interference, especially under high reporting rate conditions. 
This evaluation demonstrates that \SystemName network performs better in scenarios where tag throughput is required to be high.

\subsubsection{Impact of Readers' Density}

We deploy 7 tags and set the reporting rate of each tag to 400pps. We set the distance between each pair of readers to 1.4m, 1.8m, and 2.2m to change the reader's density and measure the overall throughput in each setting.

\begin{figure}[t]
    \centering
    \begin{minipage}[t]{0.47\linewidth}
        \includegraphics[width=\linewidth]{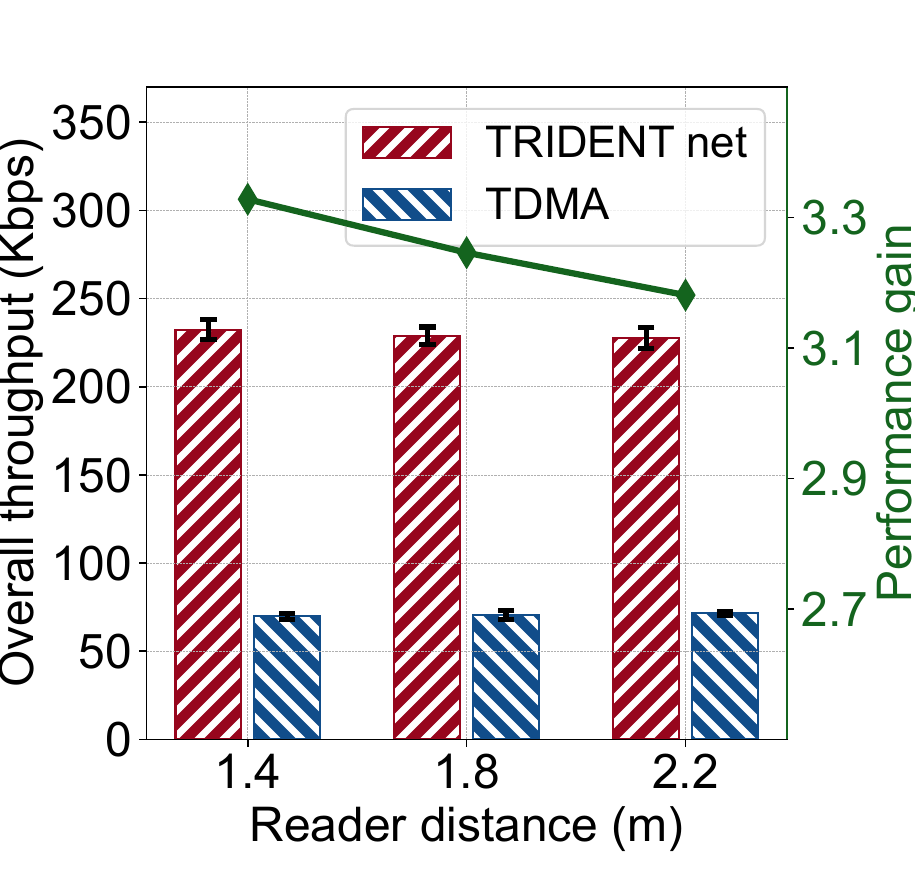}
        \caption{Overall throughput comparison between two networks with reader density represented by the distance between readers.}
        \label{fig:throughput-reader-density}
    \end{minipage}
    \hspace{0.2cm}
    \begin{minipage}[t]{0.47\linewidth}
        \includegraphics[width=\linewidth]{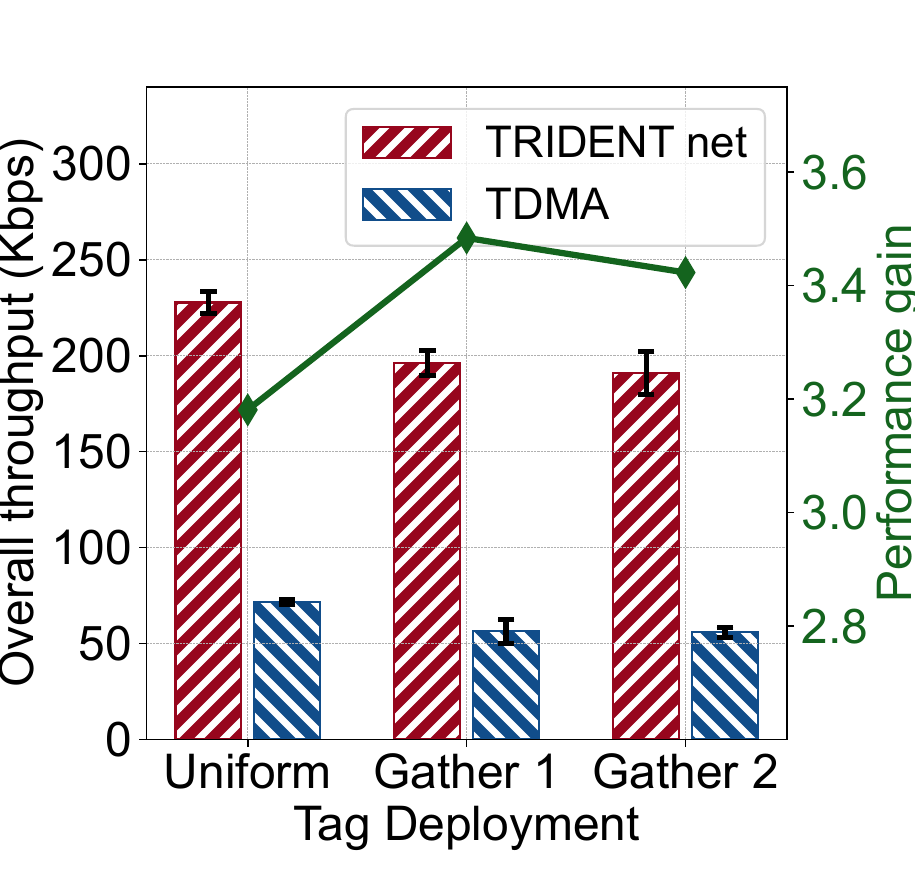}
        \caption{Overall throughput comparison with three different tag deployment methods in the network.}
        \label{fig:throughput-tag-deployment}
    \end{minipage}
    \narrowfig
    \vspace{-4mm}
\end{figure}

The results are depicted in Fig. \ref{fig:throughput-reader-density}, as the distance between readers decreases, the throughput of \SystemName remains relatively constant at 230Kbps or even slightly increases. However, the throughput of the TDMA network decreases from 72Kbps to 69Kbps. From a performance gain perspective, the closer the distance between readers, the higher the gain of \SystemName. The performance gain increases from 3.18$\times$ at a reader distance of 2.2m to 3.33$\times$ at a distance of 1.4m. This is because the TDMA network activates more tags with higher reader density, leading to more severe tag-to-tag interference and adversely affecting the throughput. The result indicates that \SystemName is more suitable for environments with dense reader deployments than the TDMA network.


\addsec
\subsubsection{Impact of Tag's Deployment}

We deploy 7 tags in the network, with each tag set to a reporting rate of 400pps. In addition to uniformly deploying tags throughout the entire experimental area, we deploy tags in the left half and front half of the area, creating two types of non-uniform tag deployments. We assess the throughput of the two networks under these three different tag deployment scenarios.

We deployed 7 tags using two additional deployment methods. In these two alternative deployments, the tags are no longer uniformly distributed. We measured the overall throughput of the two networks under these two special deployments as well as the uniform deployment setting.

In the non-uniform tag deployment environment, there may be a high concentration of tags near some readers, resulting in dense tag deployments in some areas. This uneven deployment negatively affects the throughput of both networks. As shown in Fig. \ref{fig:throughput-tag-deployment}, in the two non-uniform deployment scenarios considered in the experimental setup, the throughput of the TDMA network decreases by 22\% from 72Kbps (uniform deployment) to 56Kbps, while the throughput of \SystemName changes by 15\%, decreasing from 228Kbps (uniform deployment) to 195Kbps. The performance gain of \SystemName over the TDMA network also increases from 3.18$\times$ to 3.48$\times$ and 3.42$\times$ in these scenarios. This experiment illustrates that \SystemName has greater robustness in dealing with non-uniform tag deployments.

\color{black}


\subsection{Ablation Study}

\subsubsection{Frequency Selectivity of Reflector}

We set the reader's frequencies to 700MHz, 800MHz, and 900MHz. And we set the operating frequency bands of the tag to the three aforementioned bands, respectively. The tag is placed at a distance of 70cm from the reader. We evaluate the strength of the backscattered signal received by the reader in different bands and use their difference to characterize the frequency selectivity of the reflector. 

\begin{figure}[t]
    \begin{minipage}[t]{0.47\linewidth}
         \includegraphics[width=\linewidth]{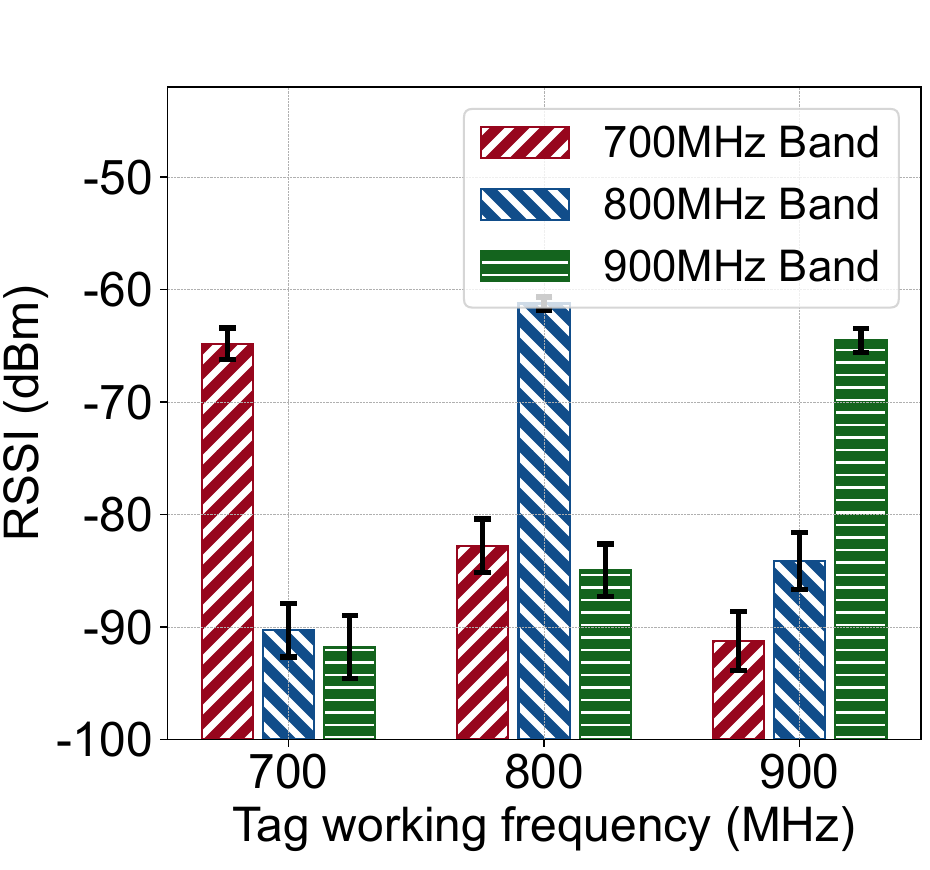}
         \caption{The RSSI of tag signals when tags operating at various frequencies are excited by different excitation frequency bands.}
         \label{fig:frequency_selectivity}
     \end{minipage}
     \hspace{0.2cm}
     \begin{minipage}[t]{0.47\linewidth}
         \includegraphics[width=\linewidth]{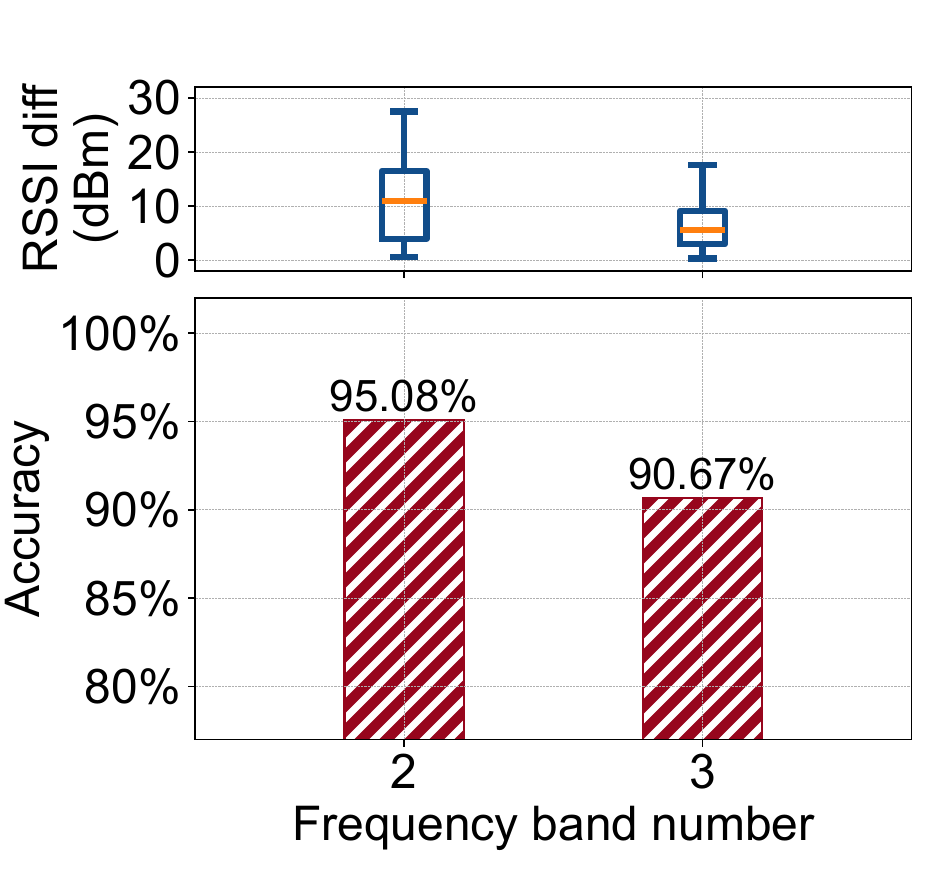}
         \caption{The accuracy of detecting the strongest frequency component when a tag is excited by signals from multiple frequency bands.}
         \label{fig:band_detection_accuracy}
    \end{minipage}
     \narrowfig
     \vspace{-3mm}
\end{figure}

The results are depicted in Fig. \ref{fig:frequency_selectivity}. We observe that when the tag backscatter incident signals across multiple frequency bands, the backscattered signal received by the reader in the tag's operating frequency band is much stronger than the signal strength outside the reflector's operating frequency band. Specifically, when the reflector is set to 700MHz, the reader receives an excitation signal at -64.8dBm at the 700MHz band, whereas the excitation signals received in the 800MHz and 900MHz bands are only -82.8dBm and -91.3dBm, respectively. The difference in signal strength between the within-band and out-of-band signals exceeds 18dB. Therefore, we can conclude that the reflector can perform reflection at a single frequency band without affecting other frequency bands, demonstrating good frequency selectivity.


\subsubsection{Accuracy of Band Detection}

We assess the accuracy of frequency band detection in environments with two and three frequency bands of excitation signals respectively.
We place the \SystemName tag at an arbitrary position and record its band detection result. Then another USRP N210 is used to measure the RSSI of the excitation signals of various frequency bands at the tag's deployment point. We measure the accuracy of the frequency band detector in selecting the strongest frequency band of the excitation signal. We conduct 80 experiments under the conditions of having two and three excitation frequency bands respectively.


The results are shown in Fig. \ref{fig:band_detection_accuracy}. When there are two different frequency bands for excitation signals, the median RSSI difference between the two bands is 10.9dB. The band detection of \SystemName achieves an accuracy of 95.08\% in detecting the strongest excitation frequency band. In the case of three frequency bands for excitation signals, the median RSSI difference between the strongest and second strongest bands is 5.6dB, and the band detection of \SystemName also achieves an accuracy of 90.67\% in detecting the strongest excitation frequency band. This evaluation illustrates that in the majority of deployment locations, the frequency band detector can identify the frequency band with the strongest excitation signal with high accuracy.


\begin{figure}[t]
    \centering
    \subfigure[The throughput of the interfered tags.]{
        \includegraphics[width=0.44\linewidth]{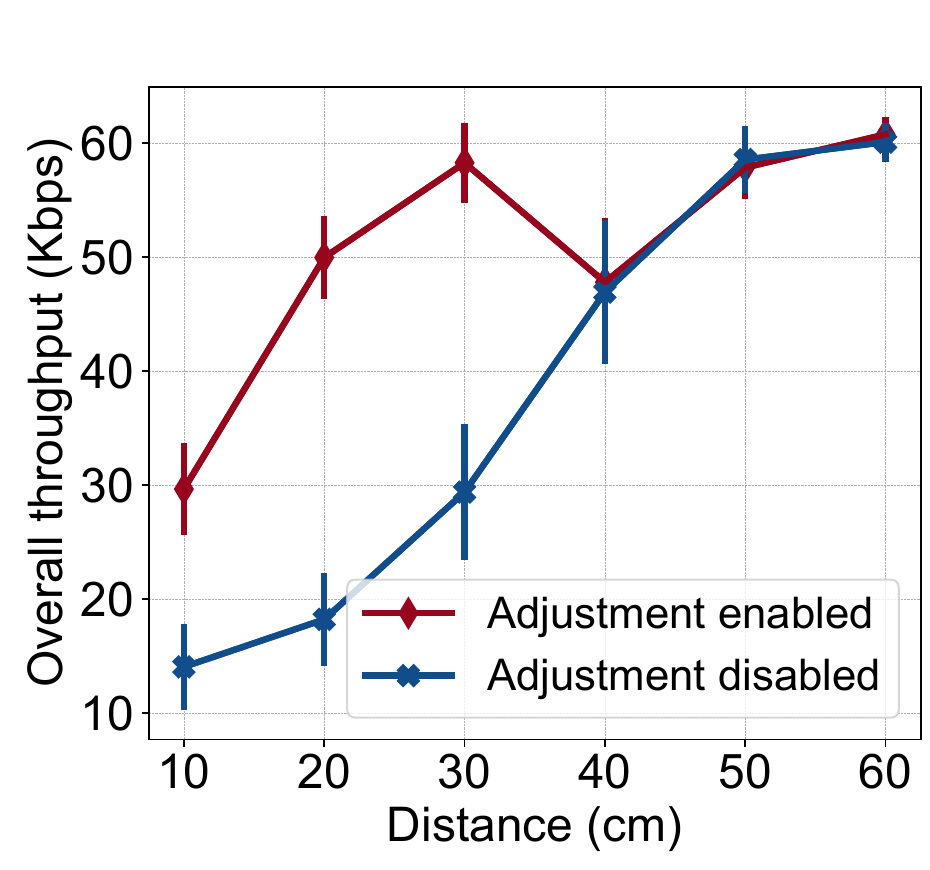}
    }
    \hspace{0.2cm}
    \subfigure[The throughput of the interfering tags.]{
        \includegraphics[width=0.44\linewidth]{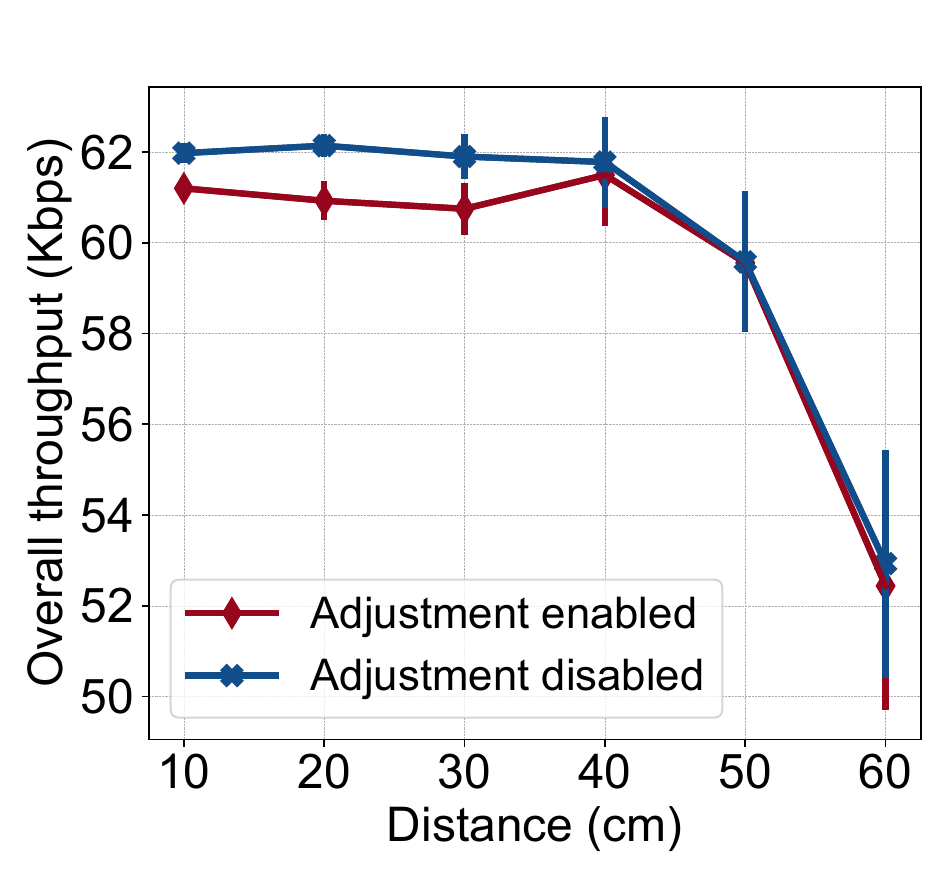}
    }
    \caption{Throughput comparison of tags with \& w/o power adjustment in different distance settings between interfering tag and readers.}
    \label{fig:power-adjustment-evaluation}
    \vspace{-5mm}
\end{figure}

\subsubsection{Throughput Increment by Adjusting Reflection Power} 
We deploy two readers, $R_A$ and $R_B$, at a distance of 3 meters and assign them to transmit excitation signals at the same frequency band. We deploy a tag $T_\alpha$ as an interfered tag at a distance of 50cm from $R_A$ and another tag $T_\beta$ as an interfering tag near $R_B$. We vary the distance $D$ between $T_\beta$ and $R_B$, and measure the throughput of both tags with reflection power adjuster enabled and disabled for the interfering tag. 

The results are shown in Fig. \ref{fig:power-adjustment-evaluation}. As the distance between $T_\beta$ and $R_B$ decreases, the throughput of the interfered $R_A$ significantly decreases due to interference. 
As the interfering tag becomes closer to the reader, the throughput of $T_\alpha$ continues decreasing if the adjuster is disabled. Especially when $D$ decreases to 10cm, the throughput of $T_\alpha$ is only 14.1Kbps. 
On the other hand, if the reflection power adjuster is enabled, the interfering $T_\beta$ with the reflection power adjuster detects excessive excitation signals and actively adjusts the reflection strength when $D$ is less than 32cm. As a result, the throughput of the interfered $T_\alpha$ increases to 58.3Kbps when $D$ is 30cm, and it can maintain a throughput of 29.6Kbps (2.1$\times$ enhancement) even when $D$ decreases to 10cm. 

Besides, the adaptive reflection power adjustment has minimal cost on the throughput of the tag $T_\beta$. When the distance $D$ is less than 30cm and the reflection power decreases, the throughput of the tag only drops from 62.0Kbps to 60.8Kbps, indicating negligible throughput loss.
\color{black}


\begin{figure}
    \centering
    \subfigure[Narrow area, algorithm]{
        \includegraphics[width=0.45\linewidth]{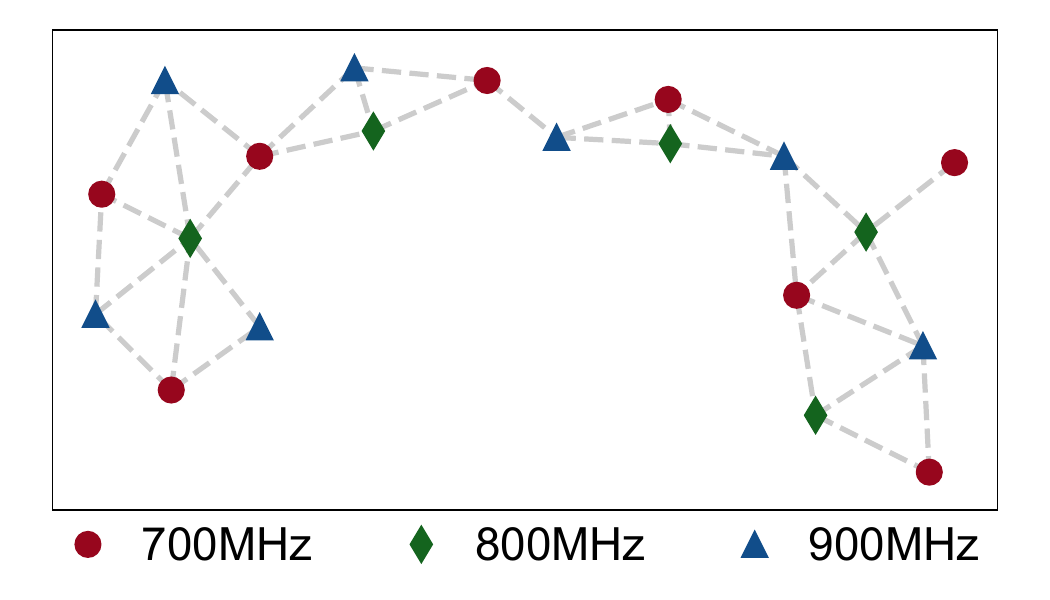}
    }
    \hspace{-2mm}
    \subfigure[Narrow area, random]{
        \includegraphics[width=0.45\linewidth]{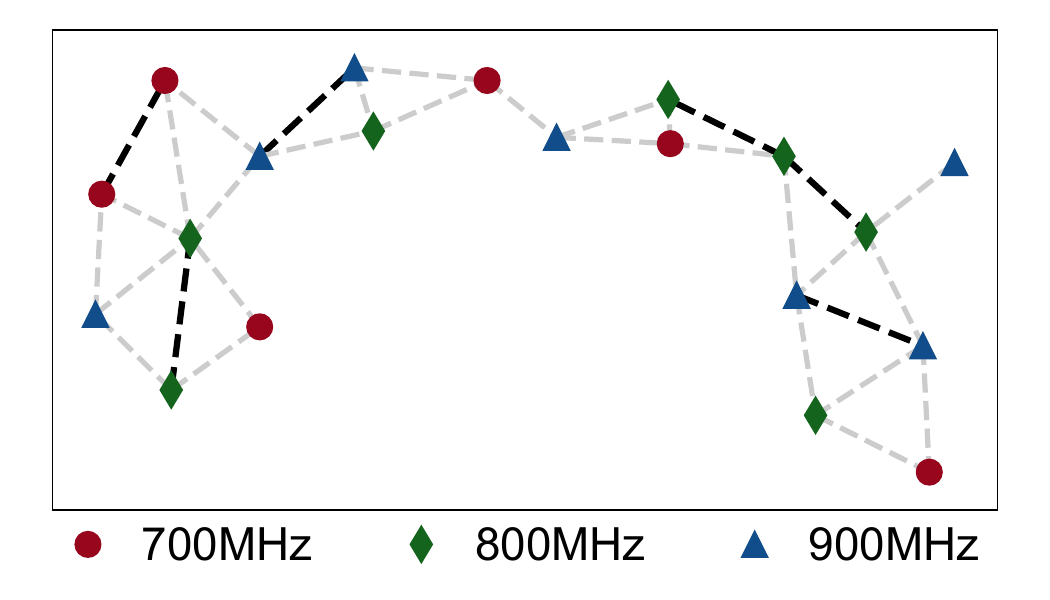}
    } 
    \quad
    \subfigure[Wide area, algorithm]{
        \includegraphics[width=0.45\linewidth]{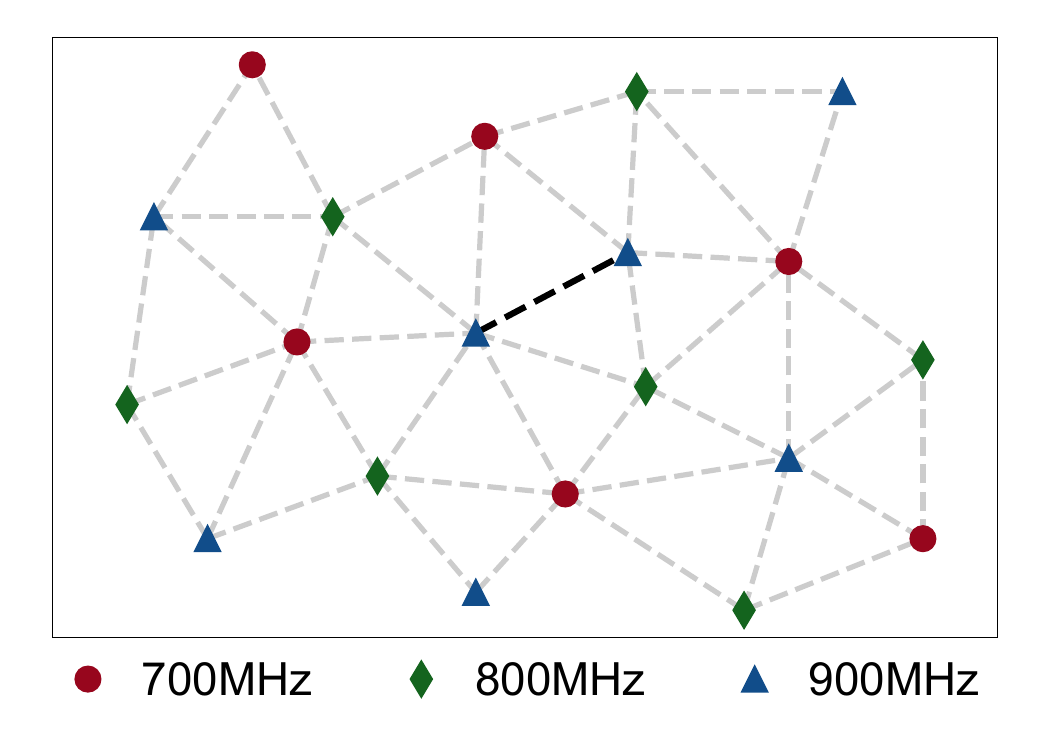}
    }
    \hspace{-2mm}
    \subfigure[Wide area, random]{
        \includegraphics[width=0.45\linewidth]{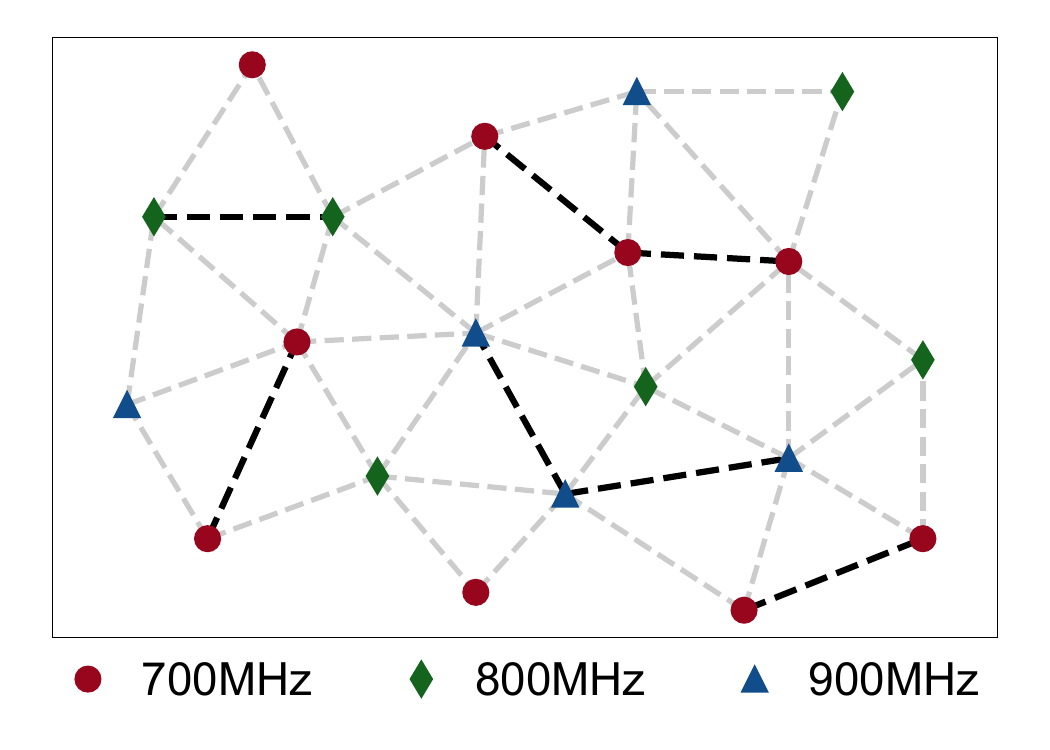}
    }       
    \caption{Interference graph and frequency assignment of our algorithm and random selection in simulation. Each point in the graph represents a reader, and each dark edge indicates potential interference.}
    \label{fig:sim-algorithm-vs-random}
    \vspace{-5mm}
\end{figure}

\subsubsection{Solution Quality of Frequency Assignment Algorithm}


We simulate wide and narrow deployment environments, each with 20 readers. We assign frequencies to the readers by the algorithm proposed in Sec.\ref{title:reader-frequency-assignment}. We set a baseline that uses the best method from 200,000 randomly generated strategies. We consider the number of adjacent reader pairs with the same frequency band as the measurement of the solution quality. 


\rev{The interference graph and simulation results are depicted in Fig. \ref{fig:sim-algorithm-vs-random}. In the narrow deployment area, our algorithm discovered a frequency allocation strategy that completely avoids interference. In the wide deployment area, our algorithm identifies a frequency assignment with only 1 pair of readers potentially causing interference. However, the baseline method exhibits more than 5 pairs of adjacent readers with the same bands and causes more severe interference.}

\begin{figure}
    \centering
    \subfigure[An example of the interference strength over
time]{
        \includegraphics[width=0.44\linewidth]{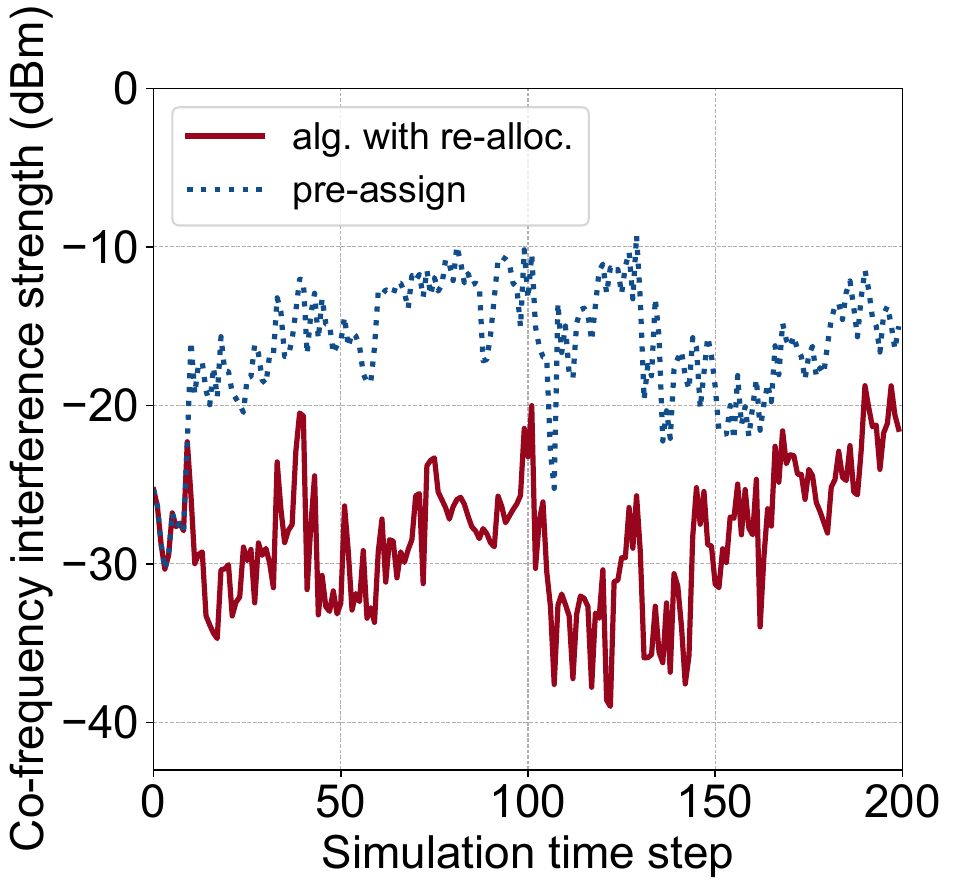}
    }
    \subfigure[CDF of the interference strength]{
        \includegraphics[width=0.44\linewidth]{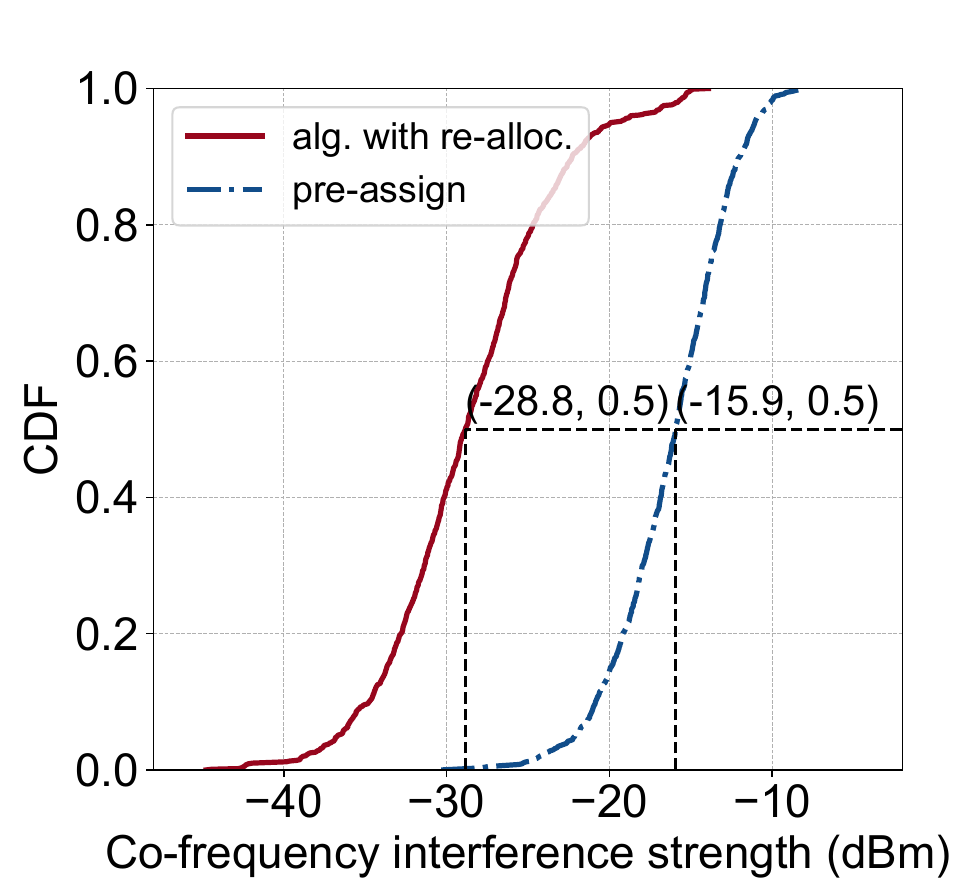}
    }
    \caption{Co-frequency interference strength of our algorithm and the pre-assigning method in the dynamic channel.}
    \label{fig:interference-in-dynamic}
    \vspace{-3mm}
\end{figure}

\subsubsection{Interference Strength in Dynamic Channel}



\rev{To further verify the performance of the algorithm in a dynamic fading wireless environment, we use a Markov stochastic process to model the fading of the wireless channel between readers \cite{seetharam2011markov}. We ran both the pre-assigning method and our frequency allocation algorithm with the reallocation mechanism in this environment, simulating the co-channel interference strength at each moment for both methods.}

\rev{We record the co-frequency interference strength at 10,000 simulation moments, plotting an example of the interference strength over time and the CDF of the co-frequency interference strength in Fig. \ref{fig:interference-in-dynamic}. We found that the median interference strength of our algorithm in the simulation is 55.3\% of that of the pre-assigned method. This is because the pre-assigned method cannot adapt to the changing channel fading, whereas our algorithm can adopt new strategies when the channel environment changes significantly.}

\color{black}

\section{Related Work}
\label{title:related-work}


Due to the highly restricted energy consumption, backscatter tags cannot utilize the coordination schemes typically applied in traditional communication systems to mitigate interference. To address the issue, solutions have been proposed from two perspectives: interference avoidance and parallel decoding.

\vspace{-3mm}
\subsection{Interference Avoidance}

The core of interference avoidance is to coordinate the signals onto orthogonal dimensions, utilizing the orthogonality to avoid interference~\cite{weiguo2023micnest, he2023acoustic}. Typically, the orthogonal spatial dimensions used by researchers include the time domain, frequency domain, code domain, and spatial domain.

\rev{There are many existing works on utilizing the time domain for interference avoidance in the multi-reader network.} These methods usually coordinate the tags backscatter at a desired time \cite{waldrop2003colorwave, birari2005mitigating, gandino2011increasing, katanbaf2021multiscatter, perez2023deepgantt, zhang2017freerider, liu2017full}. 
In \cite{gandino2011increasing, waldrop2003colorwave}, the reader runs a distributed slot reservation protocol, excites tags only in reserved time slots. 
\cite{birari2005mitigating} proposes a scheme where the reader listens to channels before exciting the tag. It will back off when it detects that another reader is working. 
In \cite{katanbaf2021multiscatter, perez2023deepgantt, zhou2007slotted}, the readers are controlled by a server, which schedules the readers' working slots. 
\rev{Some methods use multiple channels to provide more bandwidth. However, the critical problem in the multi-reader network is how to utilize the bandwidth resources. And these methods based on simple frequency division can't use the bandwidth resources effectively to avoid the interference, unless the tags are frequency-selective \cite{alesii2015multi}.}

In single-reader scenarios, some approaches point out that the frequency domain can be also utilized for backscatter tags' coordination\cite{zhao2019ofdma, zhu2020digiscatter, mitsugi2018perfectly}. 
Especially, methods proposed in \cite{zhao2019ofdma, zhu2020digiscatter} enable the OFDMA backscatter system by utilizing the frequency shifting on tags, which achieves interference avoidance in the single-reader network.
\rev{However, in the multi-reader network, these methods can't adapt to the backscatter network topology changes caused by the dynamic wireless channel, and the tags are still possible to interfere with the adjacent readers.} 
\rev{Besides, the frequency shifting requires a high-frequency clock or VCO, which causes more power consumption \cite{zhao2019ofdma, jiang2021long}}.

There have also been efforts exploring backscatter coordination in the spatial domain and code domain\cite{mi2019cbma, mutti2008cdma, mazaheri2021mmtag}. 
\cite{mi2019cbma} proposed a CDMA backscatter system by using different pseudo-noise (PN) codes on tags to spread their information. 
\cite{mazaheri2021mmtag} suggests utilizing narrow beams of the millimeter-wave to spatially coordinate backscattered signals, thus avoiding interference. 
\rev{\cite{yoon2024mmcomb} further utilizes the beamforming mechanism of commercial mmWave Wi-Fi to achieve interference avoidance. \cite{atsutse2023leakyscatter} utilizes the leaky wave antennas (LWA), to avoid interference through frequency-space division in the above-100GHz frequency band. }
Different from these works, \SystemName proposes a novel interference avoidance scheme based on frequency-space division in multi-reader environments.

\vspace{-3mm}
\subsection{Parallel Decoding}

Parallel decoding is a class of schemes that directly demodulate the information from interfered or collided signals to extract the tag's information.
In \cite{jin2017fliptracer, guo2020aloba, hu2015laissez, jin2018parallel, ou2015come}, parallel decoding algorithms are proposed. These methods utilize the symbol clusters on constellation diagrams caused by collided signals, to directly demodulate the information from collided signals. 
\rev{Specifically, these methods can handle small-scale collisions (e.g. the single-reader scenario). But in a multi-reader setting, a reader can be interfered with by tags excited by many adjacent readers. Under such complex interference, the communication and computation expense of these methods will significantly increase and eventually become unaffordable. While \SystemName utilizes the tag's frequency selectivity to avoid interference at a very low cost.}

Researchers also notice the sparsity of the FFT spectrum after dechirping LoRa chirps. They have proposed parallel demodulation algorithms that can handle up to 256 LoRa backscattered signals\cite{jiang2021long, hessar2019netscatter, peng2018plora}. 
By exploring the characteristic that the farther a tag is from the reader, the lower its reflected energy, \cite{guo2018design} utilizes power-domain Non-Orthogonal Multiple Access (NOMA) technology and receives messages from multiple tags.
\vspace{-3mm}
\section{Conclusion}
\label{title:conclusion}

We present the design, implementation, and evaluation of \SystemName, a novel tag design to enable interference avoidance based on frequency-space division. 
The tag detects channel conditions and adaptively adjusts the frequency band and power of backscattered signals.
\add{We also propose an algorithm, assigning the adjacent readers to operate at different bands.}
With these designs, the \SystemName network avoids interference while maintaining high overall throughput by utilizing both the frequency and space domain. 
The results demonstrate that \SystemName enhances the network throughput by 3.18×, compared to the TDMA-based scheme.
\section*{Acknowledgment}

This work is supported by the National Natural Science Foundation of China under grant No. U21B2007 and No. 62425207.



\bibliographystyle{IEEEtran}
\bibliography{ref.bib}

\begin{IEEEbiography}[{\includegraphics[width=1in,height=1.25in,clip,keepaspectratio]{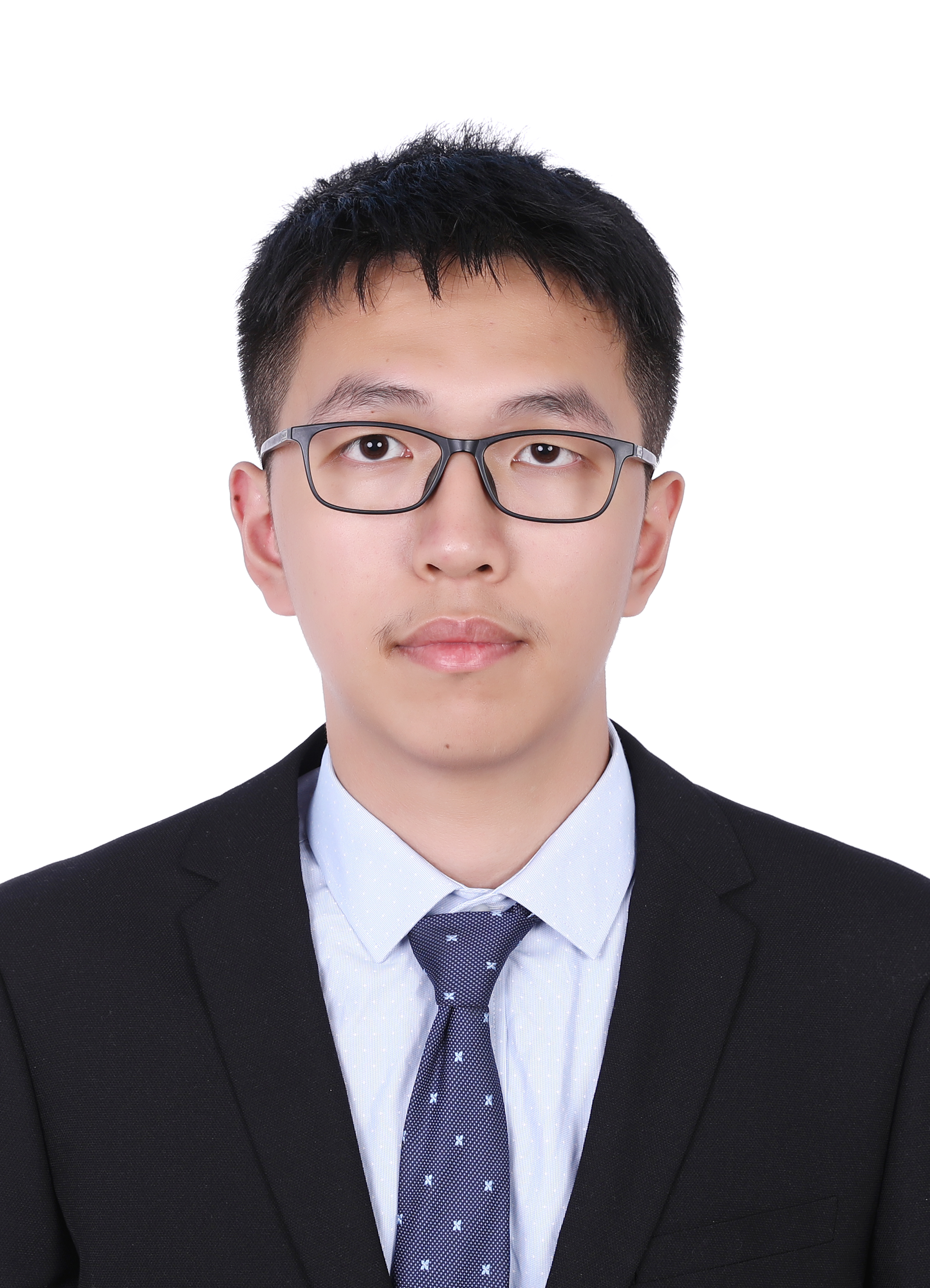}}]{Yang Zou}
is currently a PhD. student at Tsinghua University. He received his B.E. degree from the Beijing University of Aeronautics and Astronautics (BUAA). His research interests include wireless networking and communication.
\end{IEEEbiography}

\begin{IEEEbiography}[{\includegraphics[width=1in,height=1.25in,clip,keepaspectratio]{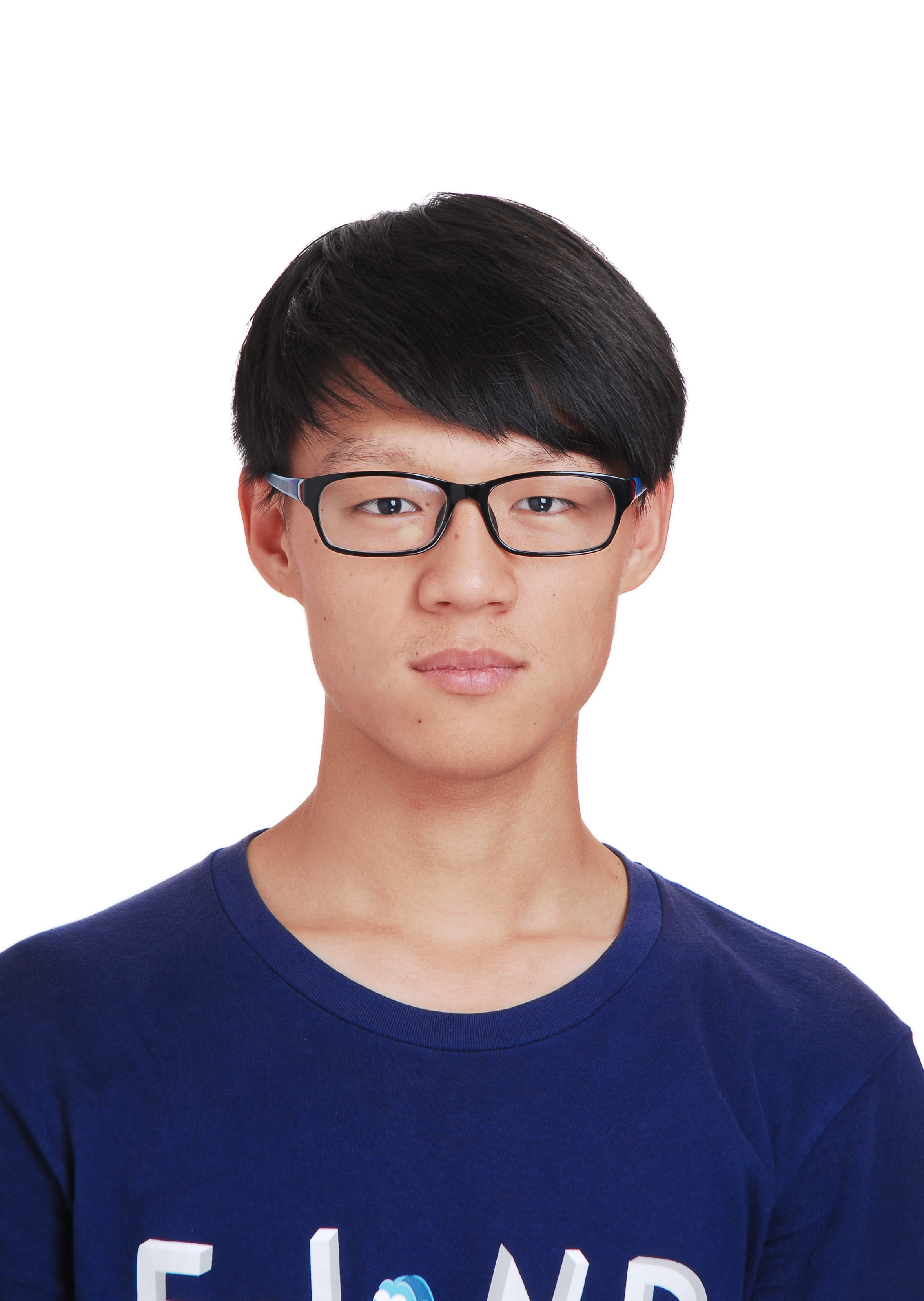}}]{Xin Na}
is currently a PhD. student at Tsinghua University. He received his B.E. degree from Tsinghua University. His research interests include wireless networking and low-power IoT.
\end{IEEEbiography}

\begin{IEEEbiography}[{\includegraphics[width=1in,height=1.25in,clip,keepaspectratio]{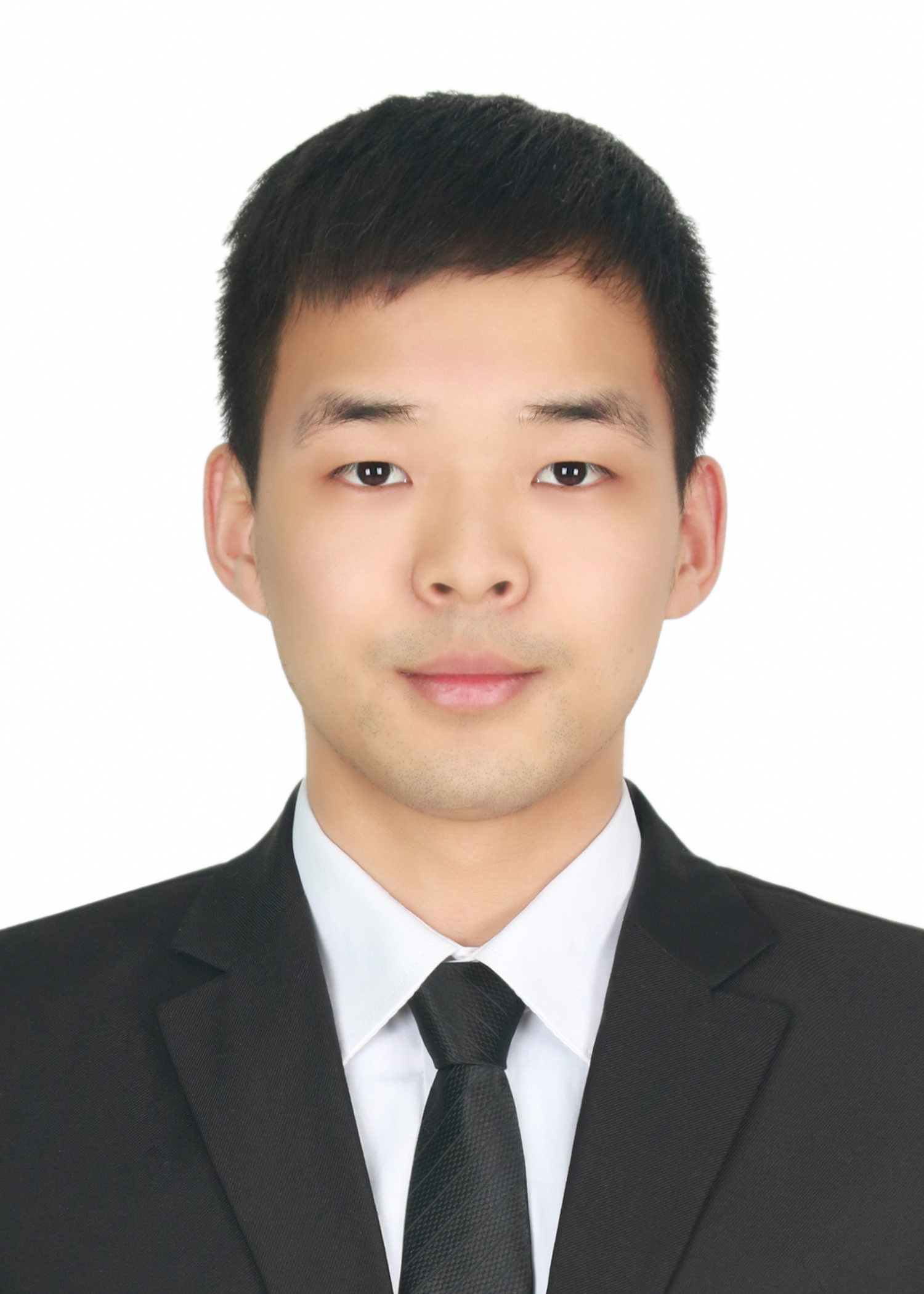}}]{Yimiao Sun}
is currently a PhD. student at Tsinghua University. He received his B.E. degree from the University of Electronic Science and Technology of China (UESTC). His research interests include mobile computing and wireless sensing.
\end{IEEEbiography}

\begin{IEEEbiography}[{\includegraphics[width=1in,height=1.25in,clip,keepaspectratio]{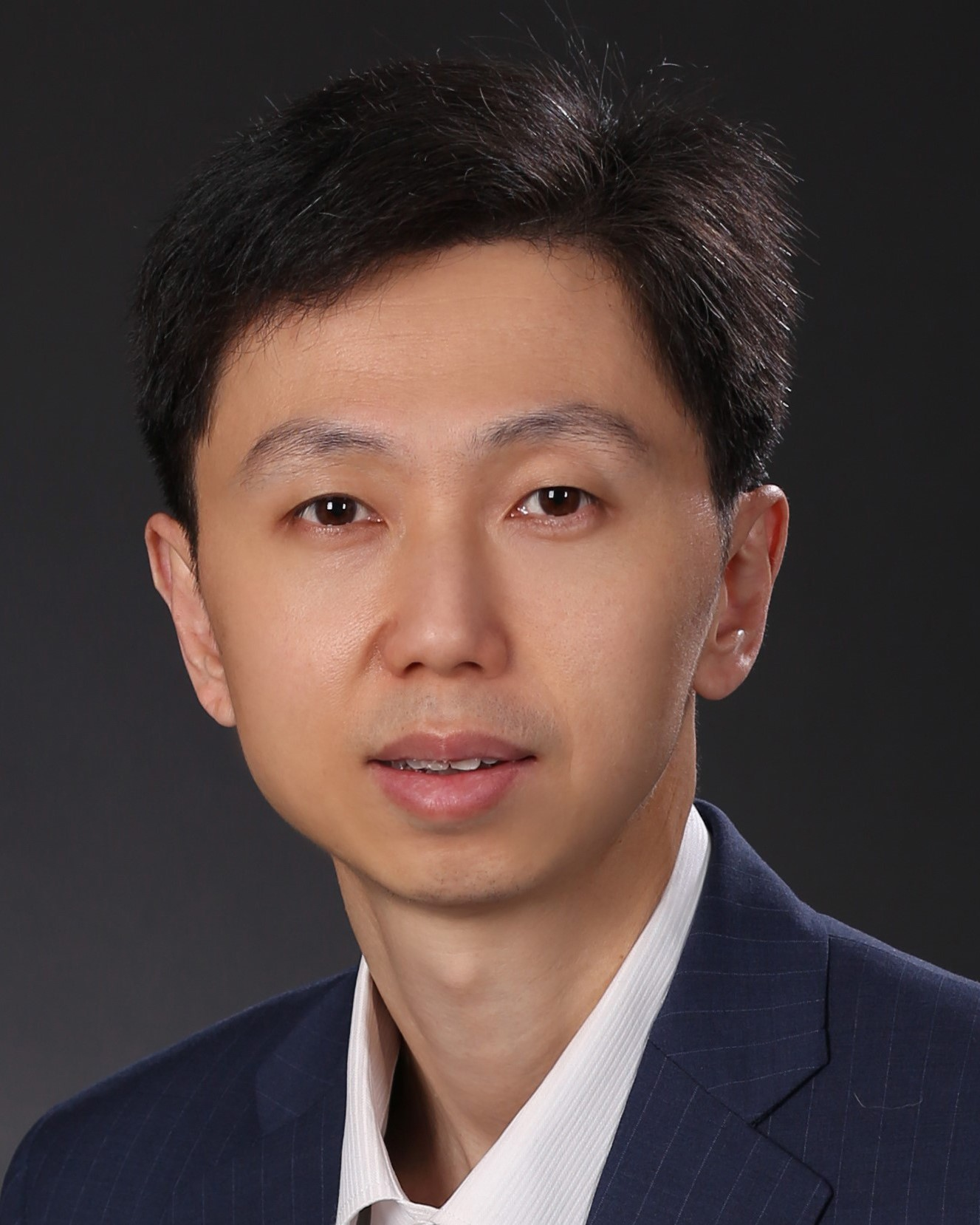}}]{Yuan He}
is an associate professor in the School of Software and BNRist of Tsinghua University. He received his B.E. degree from the University
of Science and Technology of China, his M.E. degree in the Institute of Software, Chinese Academy of Sciences, and his PhD degree in Hong Kong University of Science and Technology. His research interests include wireless networks, Internet of Things, pervasive and mobile computing. He is a senior member of IEEE and a member of ACM.
\end{IEEEbiography}

\vfill


\end{document}